\documentclass[10pt,preprint]{aastex62}
\usepackage{amsmath}
\usepackage{amssymb}
\usepackage{natbib}
\usepackage{graphicx}
\usepackage{color}
\usepackage[outdir=./]{epstopdf}
\usepackage{ulem} 
\usepackage[shadow]{todonotes} 
\usepackage{subfigure}

\newcommand{\mbf}{\boldsymbol} 
\newcommand{\I}{\mathrm{i}} 

\submitjournal{ApJ}

\shorttitle{Inference of solar rotation from perturbations of mode eigenfunctions}
\shortauthors{Schad et al.}

\begin{document}
\title{Inference of solar rotation from perturbations of acoustic mode eigenfunctions}
\correspondingauthor{Markus Roth}
\email{mroth@leibniz-kis.de}

\author[0000-0002-0786-7307]{Ariane Schad}
\affiliation{Leibniz-Institut f\"ur Sonnenphysik (KIS)\\
79104 Freiburg, Germany}
\affiliation{Freiburg Center for Data Analysis and Modeling (FDM)\\
79104 Freiburg, Germany}
\author[0000-0002-1430-7172]{Markus Roth}
\affiliation{Leibniz-Institut f\"ur Sonnenphysik (KIS)\\
79104 Freiburg, Germany}
\nocollaboration

\begin{abstract}
Today's picture of the internal solar rotation rate profile results essentially from helioseismic analyses of frequency splittings of resonant acoustic waves. Here we present another, complementary estimation of the internal solar rotation rate using the perturbation of the shape of the acoustic waves. For this purpose we extend a global helioseismic approach developed previously for the investigation of the meridional flow~\cite{schad11, schad12, schad13} to work on the components of the differential rotation. We discuss the effect of rotation on mode eigenfunctions and thereon based observables. Based on a numerical study using a simulated rotation rate profile we tailor an inversion approach and also consider the case of the presence of an additional meridional flow. This inversion approach is then applied to data from the MDI (Michelson Doppler Imager aboard the Solar Heliospheric Observator (SoHO)) instrument and the HMI (Helioseismic and Magnetic Imager aboard the Solar Dynamics Observatory (SDO)) instrument. In the end, rotation rate profiles estimated from eigenfunction perturbation and frequency splittings are compared. 
The rotation rate profiles from the two different approaches are qualitatively in good agreement, especially for the MDI data. 
Significant differences are obtained at high latitudes $> 50^{\circ}$ and near the subsurface. 
The result from HMI data shows larger discrepancies between the different methods. 
{We find that the two global helioseismic approaches} provide complementary methods for measuring the solar rotation. 
Comparing the results from different methods may help to reveal systematic influences that affect analyses based on eigenfunction perturbations, like meridional flow measurements.
\end{abstract}
\keywords{Sun: helioseismology --- Sun: observation--- Methods: data analysis}
\section{Introduction}
One of the most considerable achievements of global helioseismology is the determination of the solar rotation rate in the deep interior, e.g.,~\cite{thompson93,howe09}. 

The basis for that analysis is advection of the $p$ modes by the Sun's interior differential rotational flow, which lifts the degeneracies between the modes with the same azimuthal order $m$ within a multiplet with harmonic degree $\ell$ and radial order $n$~\citep{cowling-newin-1949, Ledoux1949, Hansen-Cox-vanHorn1977, Gough1981}. Based on forward modeling respective sensitivity functions for each individual mode $(n,l,m)$ are calculated, which relate the integral effect of the differential rotation within the propagation area of a mode to a resulting observable frequency shift. Given such frequency shifts, inversion methods based on an expansion of the rotational splittings in odd powers of $m$ were introduced \citep{Brown-et-al1989, ritzwoller91, Pijpers1997}, which allowed determining the components of the differential rotation \citep{Korzennik-et-al1988, schou98}. Other approaches besides these so-called 1.5-dimensional inversions for the interior rotation rate are those delivering results in two dimensions directly, i.e. the rotational profile as function of depth and latitude averaged over longitude~\citep{schou_et_al_1994}. These methods have in common that they deliver an estimate of the interior rotation profile averaged over the northern and southern hemisphere of the Sun (e.g.~\cite{schou94, howe09}).
The overall result is that the latitudinal differential rotation profile observed on the surface continues radially inward until it transits at the base of the convection zone (i.e. the tachocline) into rigid rotation of the solar radiative zone (see~\cite{howe09} and references therein).

In this paper we aim to present a complementary approach of estimating the solar interior rotation by investigating the influence of differential rotation on the perturbation of eigenfunctions. The formal influence of rotation on mode eigenfunctions and mode coupling was investigated by \citet{woodard89} and \citet{vorontsov11} for modes of large harmonic degree. They derived approximate expressions to quantify the strength of mode coupling due to rotation when mode coupling is restricted to modes of the same radial order.
The coupling of modes due to large-scale flows was generally discussed by ~\citet{lavely92}. Based on their approach of quasi-degenerate perturbation theory, \citet{schad11} developed a method to estimate the solar meridional flow from eigenfunction perturbations~\citep{schad12, schad13}.
In the following we will extend this method to consider the influence of solar rotation on the mode eigenfunctions. We derive observable quantities in form of mode amplitude ratios.
In the end this allows us to evaluate the mode coupling due to rotational advection. We employ data from the SoHO/MDI instrument and the {SDO/HMI} instrument for modes of low and medium harmonic degree $l<200$. We derive the coupling coefficients in terms of the toroidal flow components and present an inversion for solar rotation from the eigenfunction perturbations. 
These inversions are compared to results respectively obtained from the conventional global helioseismic approach of frequency splittings.

The paper is structured as following. In section~\ref{sec:theorymethods} we present the theoretical framework of mode coupling due to rotation.In section~3
we investigate the influence of solar rotation on mode coupling, general matrix elements, and the respective consequences on observables, especially the amplitude ratios, by means of modelled flow profiles. 
The data analysis is described in section~4.
Solar rotation rate profiles estimated from MDI and HMI data are presented in section~5.
The results are discussed in Section~6.
\section{Theory and Methods}
\label{sec:theorymethods}
We extend the theoretical framework on the mode eigenfunction perturbation analysis (EFPA) derived in~\cite{schad11,schad13}  for an axisymmetric poloidal velocity fields to an axisymmetric toroidal velocity field. 
Different reference frames are used in literature for the derivation of the equation of motions and the matrix elements, like a co-rotating accelerated frame or a non-rotating inertial frame~\citep{ritzwoller91,lavely92}. The subsequent derivations are given with respect to a heliocentric inertial frame with spherical coordinates with radius $r$, co-latitude $\theta$, and longitude $\phi$. The flow-free, non-magnetic, and non-rotating solar reference model, e.g. solar model S~\citep{dalsgaard96}, as well as its perturbations are assumed to be stationary. Any indirect perturbations of structural quantities, e.g. of density due to rotational asphericity, are neglected. The unperturbed resonant acoustic waves, $p$ modes, refer to the solar reference model. Each mode is characterized by its angular frequency $\omega_{k}$ and eigenfunction $\mbf{\xi}_{k}^0$, where the triple $k=(n,l,m)$ refers to the radial order $n$, harmonic degree $l$, and azimuthal order $m$. The eigenfunctions are orthogonal over the solar sphere and we additionally assume they are normalized, i.e., $\int \mbf{\xi}_{k'}^{0}\cdot \mbf{\xi}_{k}^{0} \,dV=\delta_{k'k}$. 
\subsection{The Toroidal Velocity Field of Solar Rotation}
Theoretically, solar rotation can be described by an axisymmetric, toroidal velocity field $\mbf{u}_{rot}(\mbf{r})=\mbf{\Omega}(\mbf{r})\times\mbf{r}$, 
where $\mbf{\Omega}(\mbf{r})$ is the angular velocity vector pointing along the axis of rotation and $\Omega(\mbf{r})$ is the rotation rate at location $\mbf{r}$. 
In spherical coordinates, the velocity field is expanded with respect to spherical harmonics $Y_{s}^{t}$ of degrees $s$~\citep{ritzwoller91}: 
\begin{eqnarray}
\label{eq:uYs1}
\mbf{u}_{rot}(r,\theta,\phi)&=&\Omega(r,\theta)\, r \sin\theta\,\mbf{e}_{\phi}\, ,\\
\label{eq:uYs2}
&=& -\sum_{s=1,2,\dots}w^{0}_{s}(r)\,\partial_{\theta}Y_{s}^{0}(\theta,\phi)\,\mbf{e}_{\phi}\, , 
\end{eqnarray}
with the azimuthal order $t=0$ due to the axisymmetry, and the toroidal expansion coefficients $\{w^{0}_{s}(r)\}$ express the radial dependency of the individual rotation rate components.
\begin{equation}
\label{eq:Omega}
\Omega(r,\theta)=-\sum_{s}\frac{w^{0}_{s}(r)}{r}\,\frac{1}{\sin\theta}\partial_{\theta}Y_{s}^{0}(\theta,\phi)\, .
\end{equation}

\subsection{Advective coupling of $p$ modes}
\label{sec:theory}
Rotation as well as any other solar velocity field $\mbf{u}$ advects the resonant acoustic waves and hence distorts both eigenfrequency and eigenfunction of the acoustic modes. If the flow speed is small compared to the speed of sound, quasi-degenerate perturbation theory can be applied and the perturbed eigenfunction $\mbf{\xi}_{k}$ of a mode $k$ is approximated by~\citep{lavely92} 
\begin{equation}
\label{eq:modecontrib}
\mbf{\xi}_{k}(\mbf{r})=\mbf{\xi}^{0}_{k}(\mbf{r})+\sum_{k'\in K_{k}\setminus\{k\}} c_{kk'} \mbf{\xi}^{0}_{k'}(\mbf{r})\, ,
\end{equation}  
where the coupling coefficients $\{c_{kk'}\}$ quantify the strength of mode coupling. Mode coupling is effective only for a subset $K_{k}\setminus\{k\}$ of modes $k'$ adjacent to mode $k$~\citep{lavely92,roth08,schad11}. The subset is specified by the quasi-degeneracy condition and rules for the coupling of angular momentum~\citep{lavely92,roth99,schad11}. The coupling coefficients $c_{kk'}$ can be further expanded by a non-degenerate perturbation approach. Up to second order, one finds~\citep{schad11,schad13a} \\
\begin{equation}
\label{eq:ceq}
c_{kk'}\approx\delta_{kk'}+(1-\delta_{kk'})\Bigg\{\frac{H_{k'\,k}}{\omega^{2}_{k}-\omega^{2}_{k'}}+\sum_{j\neq k}\frac{H_{k'j}H_{jk}}{(\omega^{2}_{k}-\omega^{2}_{k'})(\omega^{2}_{k}-\omega^{2}_{j})}-\frac{H_{k'k}H_{kk}}{(\omega_{k}^{2}-\omega_{k'}^{2})^{2}}\Bigg\}\, ,
\end{equation}
where 
\begin{equation}
\label{eq:gmelementrot0}
H_{k'k}:=2\,\I\, \omega_{ref}\int \rho_{0}\, \overline{\mbf{\xi}_{k'}} \cdot (\mbf{u}\cdot \nabla\mbf{\xi}_{k})\,d^{3}\mbf{r}
\end{equation}
is the general matrix element~\citep{lavely92} of advection. Here, $\omega_{ref}$ is a reference frequency chosen near the mode eigenfrequencies. 
In~\cite{schad11} we introduced the {\em coupling ratios} 
\begin{equation}
C_{kk'}=c_{kk'}/c_{kk}
\end{equation}
as a measure of the distortion of the eigenfunction. This normalization of the expansion coefficients is independent of the chosen normalization of the expansion coefficients.
\subsection{General matrix element of rotation}
Inserting Equation~\eqref{eq:uYs2} in Equation~\eqref{eq:gmelementrot0}, the general matrix element for mode coupling by rotation is 
\begin{equation}
\label{eq:gmelementrot}
H^{(rot)}_{k'k}=\delta_{mm'} \,\omega_{ref}\sum_{s=0}^{\infty}\Big[(-1)^{-m'}
 \begin{pmatrix}
     l' &  s & l  \\
     -m' & 0  & m\\
\end{pmatrix}
 \int_{0}^{R}\rho_{0}(r)T^{k'k}_{s}(r)\,w^{0}_{s}(r)\,r^{2}dr\Big]\, ,
\end{equation}
where $T^{k'k}_{s}(r)$ is the toroidal flow kernel~\citep{lavely92} given in Appendix~\ref{eq:toroidalkernel}.

The azimuthal dependency of the matrix elements can be expressed by the Wigner 3-j polynomials $\{\mathcal{P}^{s}_{l'l}(m)\}$~\citep{schad11,schad13} to
\begin{align}
\label{eq:Hrotacoeff}
H^{(rot)}_{n'l',nl}(m)=\omega_{ref}\sum_{s}a^{s}_{k'k}\mathcal{P}_{l'l}^{s}(m)\, ,
\end{align}
with expansion coefficients
\begin{equation}
\label{eq:acoeff}
a^{s}_{k'k}:=\int_{0}^{R}\rho_{0}(r)T^{k'k}_{s}(r)\,w^{0}_{s}(r)\,r^{2}dr\, ,
\end{equation}
which we name in the subsequent part as $a$-coefficients. 

Formally, the general matrix element of rotation equals the general matrix element of the meridional flow $H^{(merid.)}_{k'k}$, \cite[Eq. 22]{schad11}. However, for differential rotation it is purely real valued, $H^{(rot)}_{k'k} \in \mathbb{R}$, while $H^{(merid.)}_{k'k}$ is purely imaginary. Moreover, $H^{(rot)}_{k'k}(m)$ is antisymmetric with respect to $m$ since it is non-vanishing only if $(l'+s+l)$ is odd, which expresses the toroidal nature of the flow. That is in contrast to $H^{(merid.)}_{k'k}$, which is symmetric with respect to $m$, which expresses the poloidal nature of the flow.
\subsection{Self- and cross-coupling}
The general matrix elements specify different kinds of mode coupling: \textit{self-coupling} and \textit{cross-coupling}. The self-coupling of modes is represented by the diagonal matrix elements $H^{(rot)}_{kk}(m)$. They are essentially responsible for the frequency splitting of azimuthally degenerate $p$ modes (see Sec.~\ref{subsec:splitt} below). The cross-coupling of modes is determined by the off-diagonal elements $H^{(rot)}_{k'k}$ with $k'\neq k$. In the case of $t=0$ cross-coupling is possible only between modes of equal azimuthal order. 
Mode coupling is composed of \textit{direct} and \textit{indirect} coupling. Given a reference mode $k$, first order terms in Eq.~\eqref{eq:ceq} are proportional to $H_{k'k}$ and contribute to the direct cross-coupling. However, higher order terms with $H_{k'j} H_{jk}$ in the perturbation expansion contribute to the indirect cross-coupling of mode $k$ with mode $k'$ since its contribution is mediated by the coupling of both modes $k$ and $k'$ with modes $j\neq k,k'$. 

The diagonal general matrix elements of self-coupling $H_{kk}$ contribute only in second or higher order to the coupling ratios, see also Eqs.~\eqref{eq:modecontrib} and \eqref{eq:ceq}. 

The component $s=1$, specifies the uniform rotation rate in $\theta$, or rigid rotation rate. 
It is $T_{1}^{k'k}=0$ for all $k \neq k'$, i.e. the component $w_{1}(r)$ contributes to the diagonal elements of the perturbation matrix only. Therefore, the eigenfunction perturbations up to first order are not sensitive to the $s=1$-component since they are determined by the off-diagonal matrix elements of the general matrix $\mbf{H}$. As a consequence, it is not possible to deduce the complete rotation rate from a first order perturbation analysis of the mode eigenfunctions.

\subsection{Combined Effect of Differential Rotation and Meridional Flow on Mode Coupling}
If both, meridional flow and solar rotation are taken into account, the coupling coefficient for a mode $k=(n,l,m)$ and $k'=(n',l',m)$ is 
\begin{align}
c_{k'k}=c^{(rot)}_{k'k}+c_{k'k}^{(merid.)}\, , \quad k'\neq k \,,
\end{align}
where $c^{(rot)}_{k'k}\in \mathbb{R}$ is the coupling coefficient due to rotation and $c_{k'k}^{(merid.)}\in \I \mathbb{R}$ is the coupling coefficient due to meridional flow as defined in~\cite[Eq. 4]{schad13}. 

Following Eq.~\eqref{eq:Hrotacoeff} and~\citep{schad11}, the coupling coefficient up to first order approximation expressed in terms of the Wigner 3-j polynomials is
\begin{align}
\label{eq:crotmerid}
c_{k'k}\approx\frac{\omega_{k}}{{\omega_{k}^{2}-\omega_{k'}^{2}}}\sum_{s}(a^{s}_{k'k}+\I\,b^{s}_{k'k})\mathcal{P}^{s}_{l'l}(m)\, .
\end{align}

The expansion coefficients $b^{s}_{k'k}$ defined in~\citep[Eq. 5]{schad13} and $a^{s}_{k'k}$, see Eq.~\eqref{eq:acoeff}, are both real valued. 

As noted earlier, due to the axial symmetry, mode coupling induced by the meridional flow as well as differential rotation are restricted to modes of identical azimuthal order $m$. Thus, the overall coupling matrix of the modes, i.e. the supermatrix $\mbf{Z}$~\citep{lavely92}, can be arranged into a block diagonal matrix, where each block $\mbf{Z}_{m}$ is composed of
\begin{align}
\label{eq:Zallflows}
\mbf{Z}_{m}=\mbf{D}_{m}+\mbf{H}_{m}^{(merid.)}+\mbf{H}^{(rot)}_{m}\, .
\end{align}
Here, $\mbf{D}_{m}$ is a diagonal matrix with elements $\lambda_{k'k'} = \omega_{k'}^{2}-\omega_{ref}^{2}$ and $\mbf{H}_{m}^{(merid.)}$ is the general matrix of the meridional flow, as defined in~\cite[Eq.~21]{schad11}, which has only off-diagonal entries and $\mbf{H}^{(rot)}_{m}$ is the general matrix for the differential rotation, see Eq.~\eqref{eq:Hrotacoeff}. 
\subsection{Splitting coefficients}
\label{subsec:splitt}
The conventional global helioseismic approach for inferences on the interior rotation rate uses the splitting of $p$-mode frequencies. Rotation lifts the degeneracy of the eigenfrequencies $\{\omega_{nl}\}$ of modes of the same multiplet with respect to azimuthal order $m$ such that $\omega_{nlm}=\omega_{nl}+\delta \omega_{nlm}$. The rotational frequency shift expanded up to first order is~\citep{ritzwoller91}
\begin{equation}
\label{eq:wnlm}
\delta \omega_{nlm}\approx H_{kk}/2\omega_{nl0}=\sum_{s\,\text{odd}}\alpha_{s,nl}P_{sl}(m)\, ,
\end{equation}
where $\{\alpha_{s,nl}\}$ represent Clebsch-Gordon splitting coefficients 
\begin{equation}
\alpha_{s,nl}:=\int_{0}^{R}\rho_{0}(r)K_{nl}^{s}(r)w^{0}_{s}(r)r^{2}dr
\end{equation}
with
\begin{equation}
\label{eq:torkernel}
K_{nl}^{s}(r)=-\frac{1}{r}((\xi_{nl}^r)^{2}+l(l+1)(\xi_{nl}^h)^{2}-[2\xi_{nl}^r\xi_{nl}^h+\frac{1}{2}s(s+1)(\xi_{nl}^h)^{2}])\, 
\end{equation}
defining the toroidal flow kernel, where $\xi_{nl}^r$ and $\xi_{nl}^h$ are the radial and horizontal components of the mode eigenfunction, resp., and 
\begin{equation}
P_{sl}(m):=-(2l+1)l(l+1)\gamma_{s}
\begin{pmatrix}
     l & s &  l  \\
     -1 & 0 & 1 \\
\end{pmatrix}\, \mathcal{P}_{ll}^{s}(m)\, 
\end{equation}
are orthogonal polynomials~\citep{ritzwoller91} defined analogously to the Wigner-3j polynomials~(see e.g.~\cite{schad11}). 

Note that the sum index $s$ in Eq.~\eqref{eq:wnlm} is restricted to odd integers, since $K_{nl}^{s}(r)=0$ for $s$ even. As a consequence, splitting coefficients are not sensitive to toroidal flow components that are anti-symmetric with respect to the equator.

The presented expansion of mode eigenfunction perturbations can be considered as a generalization of the expansion in~\cite{ritzwoller91} used to describe frequency splittings due to the self-coupling of modes. In the case of self-coupling, the matrix elements and polynomial expansion equals the descriptions in~\cite{ritzwoller91}. 
\subsection{Measurement of Mode Coupling}
Mode coupling leads to crosstalk between the global oscillations of a reference mode to neighboring degrees $l$. The amount of crosstalk can be measured with the Fourier amplitude ratio $y_{lm\,l'm}(\omega_{nlm}):=\frac{\tilde{o}_{l'm}(\omega_{nlm})}{\tilde{o}_{lm}(\omega_{nlm})}$, 
of spherical harmonic transformed global oscillations $\tilde{o}_{lm}$ and $\tilde{o}_{l'm}$ evaluated at the mode frequency $\omega_{nlm}$~\citep{schad11,schad13}. The amplitude ratio is related in first order to the coupling coefficients by~\citep{schad11} 
\begin{align}
\label{eq:aratio}
y_{lm\,l'm}(\omega_{nlm}):=\frac{\tilde{o}_{l'm}(\omega_{nlm})}{\tilde{o}_{lm}(\omega_{nlm})}\approx\frac{\sum_{k''\in K_{k}} c_{kk''}L_{k'k''} \xi^{r}_{k''}(R)}{\sum_{k''\in K_{k}} c_{kk''}L_{kk''}\xi^{r}_{k''}(R)} \in \mathbb{C}\, ,
\end{align}
where $c_{kk}=1$ and $\xi^{r}_{k''}(R)$ is the radial eigenfunction of mode $k''$ at the observation point $R$. The matrix elements $\{L_{k'k''}\}$ denote the systematic leakage of spectral power of a mode $k$ to nearby modes $k'$ induced by the limitations in the observations of acoustic waves, such as the line-of-sight projection of the velocity field and the restricted field of view to the solar front. Hence, the spherical harmonic decomposed time series do not perfectly separate even in the absence of mode coupling~\citep{schou94,korzennik04}. 

The estimation of the expected value of the amplitude ratio, i.e. the \textit{complex gain}, and the determination of the estimation error are described in~\citep[Sec. 2.2]{schad13}. 

\section{Modelling study}
\label{diffmodel}
\begin{figure}[htp]
\begin{center}
\includegraphics[width=15cm]{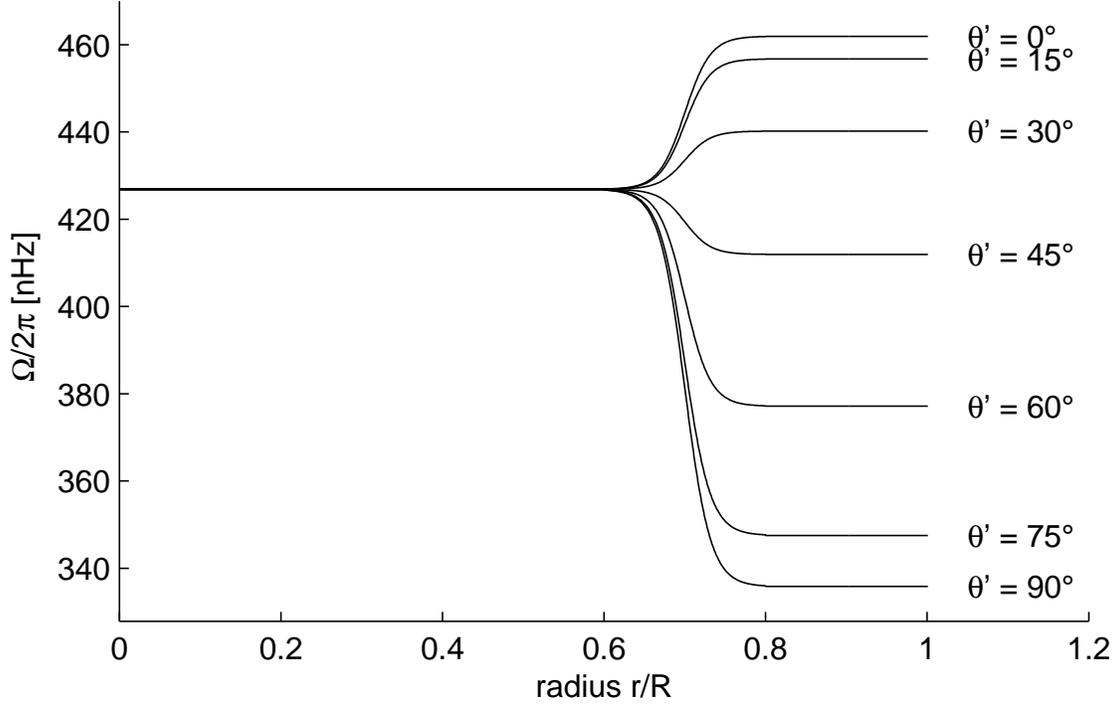}
\caption{Radial profile of the rotation rate model $\Omega(r,\theta)$ in Eq.~\ref{eq:diffrotmodel} at different latitudes $\theta'=90^\circ-\theta$.}
\label{fig:rotprofiles_simu}
\end{center}
\end{figure}

We investigate the properties and magnitude of $H^{(rot)}_{k'k}(m)$ in a theoretical modelling study. For the rotation rate we use the model 
\begin{align}
\label{eq:Omegartheta}
\Omega(r,\theta)=\sum_{s=0,2,4}\Omega_{s}(r)P_{s}(\cos\theta)\, ,
\end{align}
where $P_{s}(\cos\theta)$ is the Legendre polynomial of degree $s$ and \\
\resizebox{\linewidth}{!}{
  \begin{minipage}{\linewidth}
\begin{align}
\label{eq:diffrotmodel}
\Omega_{s}(r):=\left\{
\begin{array}{ll}
\Omega_{i}\delta_{s0} & \mathrm{if}\quad 0\leq r<r_{l}\\
1/2 \left[(\Omega_{s}(R)-\Omega_{i}\delta_{s0})\tanh(\gamma(r))+\Omega_{i}\delta_{s0}+\Omega_{s}(R)\right] & \mathrm{if}\quad r_{l} \leq r \leq r_{u}\\
\Omega_{s}(R) & \mathrm{if}\quad r_{u}<r\leq R \, ,
\end{array}\right .
\end{align}
 \end{minipage}
}\\
with $\gamma(r):=\pi(2\frac{r-r_{l}}{r_{u}-r_{l}}-1)$. This model resembles the characteristic properties of the rotation profile obtained in global helioseismology. The summation index $s$ is restricted to even degrees $s=0,2,4$ since the velocity field is assumed to be symmetric about the equatorial plane. The radii $r_{l}$, $r_{u}$ limit the region of the tachocline and are set to: $r_{l}=0.6$\,R, $r_{u}=0.8$\,R. Within the tachocline region, the model has a smooth transition from uniform rotation in the solar interior to differential rotation in the convection zone. The parameters of the rotation rate above the tachocline are chosen such that they correspond to measurements of the solar rotation rate at the surface~\citep{snodgrass83,stenflo89}: $\Omega_{0}(\mathrm{R})=2.682\,\mu$Hz, $\Omega_{2}(\mathrm{R})=-0.497\,\mu$Hz, and $\Omega_{4}(\mathrm{R})=-0.075\,\mu$Hz. The chosen parameters for the rotation rate in the solar interior are: $\Omega_{i}=2.682\,\mu$Hz.  
Profiles of this rotation rate model are displayed in Fig.~\ref{fig:rotprofiles_simu} as a function of $r$ for a selected set of latitudes $\theta'=90^\circ-\theta$. The rotational shear observed in the solar subsurface layer~(e.g.,~\cite{howe09}) is neglected in the model.
The explicit relations between $w^{0}_{s}(r)$ and $\Omega_{s}(r)$ are given in App.~\ref{app:rotationws}. This model for the rotation rate given in Eq.~\eqref{eq:diffrotmodel} together with~\eqref{eq:w1}--~\eqref{eq:w3} is used for computing general matrix elements $H_{k'k}^{(rot)}(m)$ according to Eq.~\eqref {eq:gmelementrot} using eigenfunctions and eigenfrequencies of Model S for modes with $0\leq l\leq 200$.

\begin{figure}[htp]
\begin{center}
\includegraphics[width=15cm]{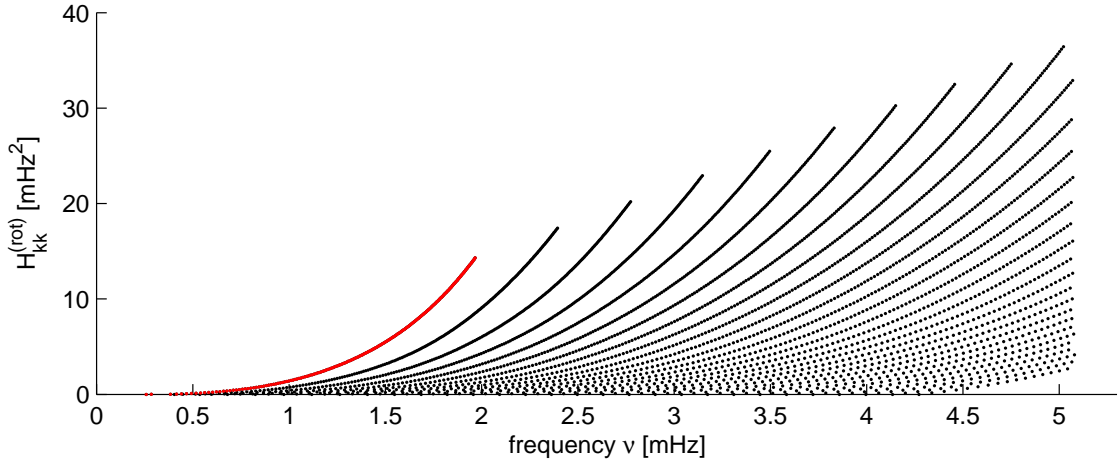}
\caption{General matrix elements ${H_{kk}}^{(rot)}(m)$ for self-coupling due to differential rotation as a function of mode frequency for modes with azimuthal order $m=l$ and harmonic degrees $0\leq l\leq 200$. Red dots highlight modes with radial order $n=1$.}
\label{fig:Hrotkk}
\end{center}
\end{figure}

\begin{figure}[htp]
\begin{center}
\includegraphics[width=15cm]{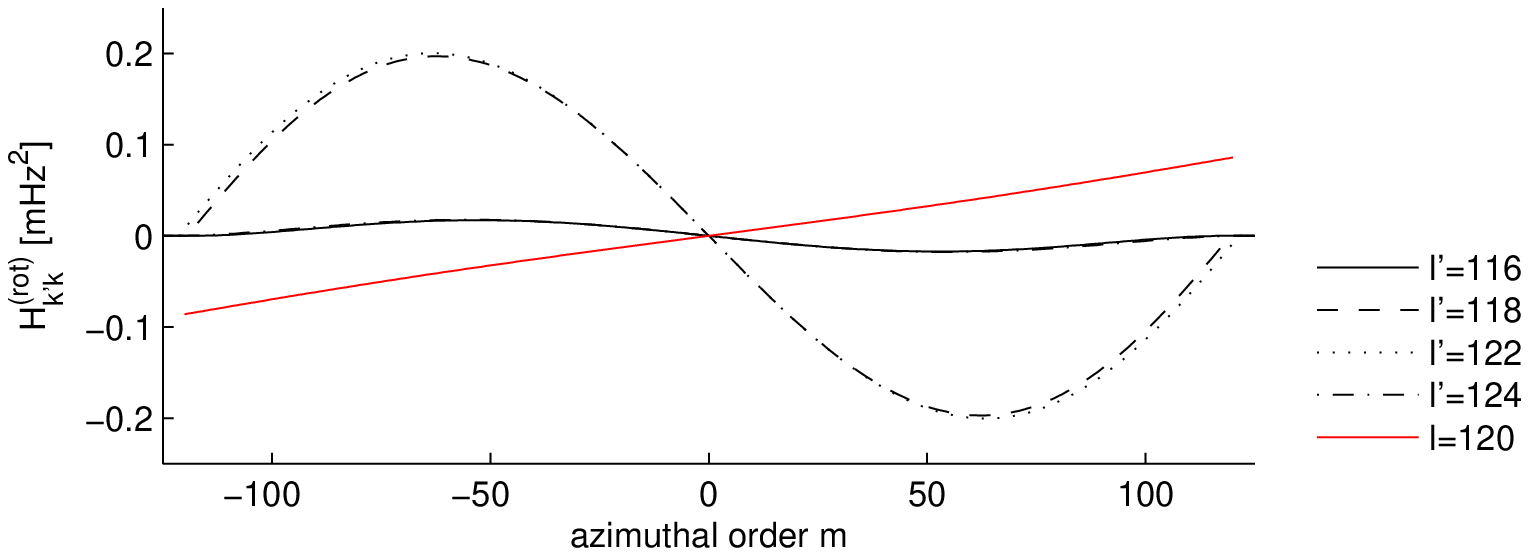}
\caption{The general matrix elements $H^{(rot)}_{k'k}(m)$ as a function of azimuthal order for the cases of self-coupling (red) and cross-coupling (black) due to differential rotation for the reference modes $k=(n=2,l=120,m)$ and coupling modes $k'=(n'=2,l',m)$ with $l'=116,118,122,124$. For better visibility, the matrix elements $H^{(rot)}_{kk}$ for self-coupling are scaled by a factor 1/100.}
\label{fig:Hrotkpk}
\end{center}
\end{figure}

The resulting matrix elements for self-coupling are displayed in Fig.~\ref{fig:Hrotkk} for $m=l$ as a function of mode frequency $\omega_{nl}$.  The matrix elements are aligned along ridges of equal radial order $n$ and increase with increasing frequency. 
The antisymmetric azimuthal behaviour of $H^{(rot)}_{k'k}(m)$ for self-coupling and cross-coupling modes is illustrated in Fig.~\ref{fig:Hrotkpk} for the reference mode $k=(n=2,l=120)$ coupling to modes $k'$ with $n'=2$ and $l'=116,118,122,124$. In case of cross-coupling, the matrix elements are largest for median azimuthal orders $m\approx l/2$ and vanish for $m=l$. In the case of self-coupling, the matrix elements are largest for $m=l$. The results show that the magnitude of the matrix elements due to cross-coupling are typically much smaller, about a factor 100, compared to the magnitude of matrix elements due to self-coupling.   

\subsection{Coupling ratios and amplitude ratios in the presence of meridional flow and rotation}
\label{subsec:cratarat}
\begin{figure}[htp]
\begin{center}
\subfigure[$l'=116$]{\includegraphics[width=17cm]{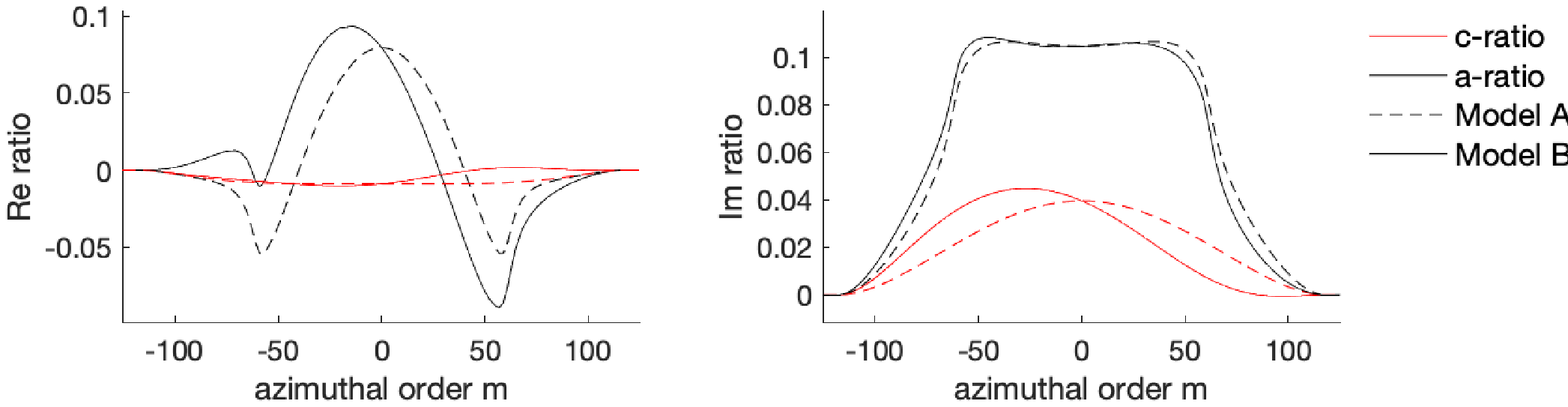}}
\subfigure[$l'=117$]{\includegraphics[width=17cm]{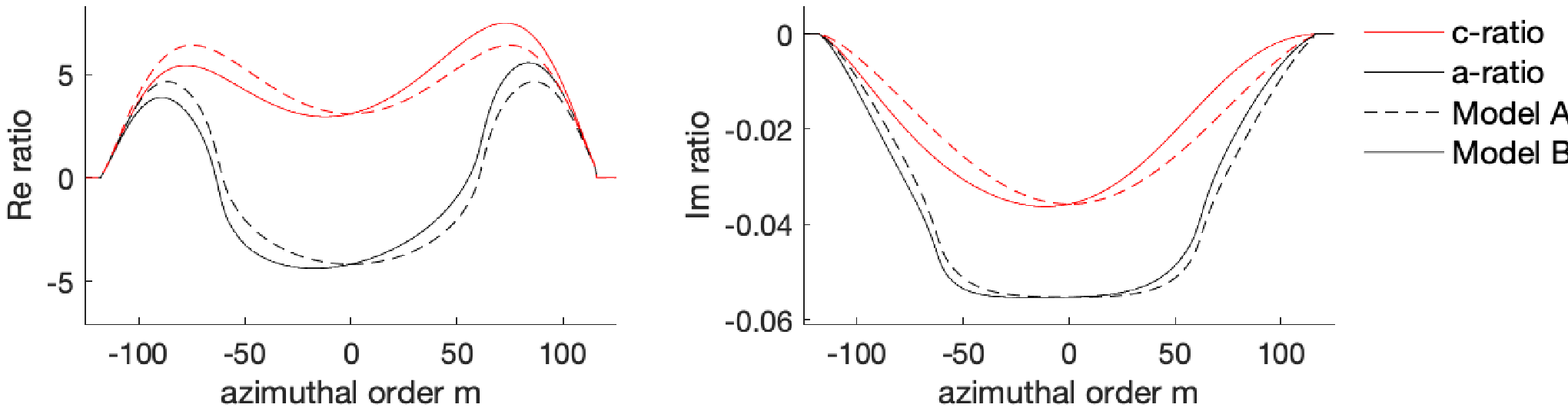}}
\subfigure[$l'=118$]{\includegraphics[width=17cm]{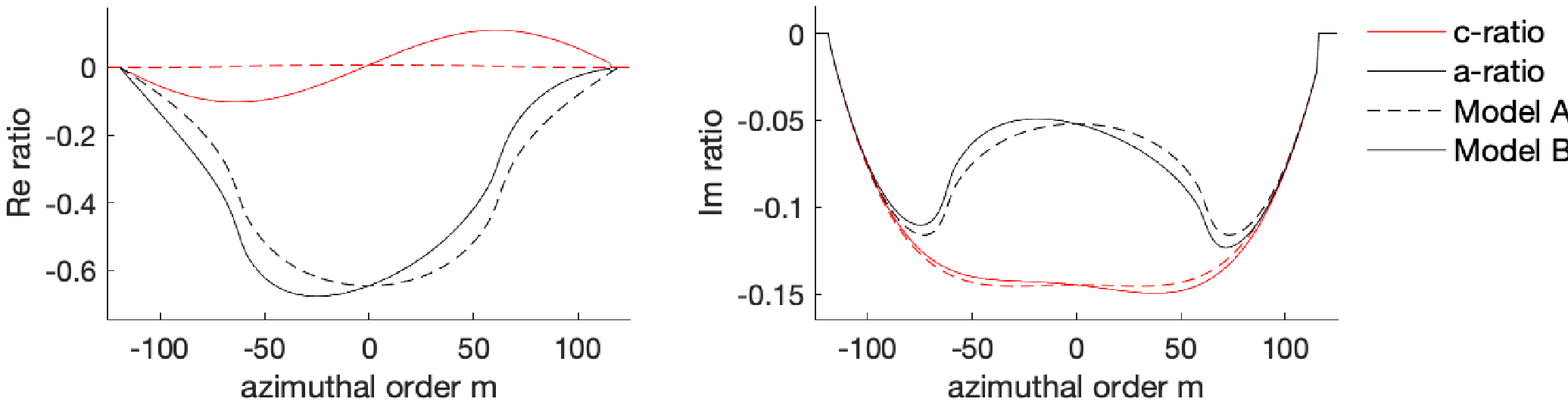}}
\caption{Coupling ratios ({red}) and amplitude ratios (black) as a function of azimuthal order $m$ for the perturbation models A and B for the reference mode $k=(n=2,l=120)$ and coupling modes $k'=(n'=2,l')$ with: a) $l'=116$, b) $l'=117$ and, c) $l'=118$. The real (imaginary) part of the ratios are shown in the left (right) column.  Model A consists of a multi-cellular meridional flow (dashed line); Model B is a superposition of the multi-cellular meridional flow with the differential rotation model in Fig.~\ref{fig:rotprofiles_simu} (continuous line).}
\label{fig:aratiosallflows}
\end{center}
\end{figure}

We investigate the influence of both rotation and meridional flow together on the coupling and amplitude ratios. Making use of the rotation model for $\mbf{u}_{rot}$ in Eq.~\ref{eq:Omegartheta} and a multi-cellular meridional flow model $\mbf{u}_{m}$ defined in Appendix~C, 
we calculated the eigenvectors of $\mbf{Z}_{m}$ in Eq.~\eqref{eq:Zallflows} for the two cases: 
\begin{description}
\item[Perturbation model A] $\mbf{u}_{m}\neq 0$ and $\mbf{u}_{rot}=0$, i.e., $\mbf{H}^{(rot)}_{m}=\mbf{0}$
\item[Perturbation model B] $\mbf{u}_{m}\neq 0$ and $\mbf{u}_{rot}\neq0$, i.e., $\mbf{H}^{(rot)}_{m}\neq \mbf{0}$.
\end{description}
The eigenvectors obtained for the two different perturbation models, resp., are further used to determine the coupling ratios and amplitude ratios by taking the leakage matrix of MDI as described in Sec.~2.7 into account.
Examples of calculated coupling ratios and amplitude ratios for the reference mode $(n=2, l=120)$ are given in Fig.~\ref{fig:aratiosallflows}.  

Comparing the results obtained from perturbation model~A and model~B in Fig.~\ref{fig:aratiosallflows} allows to conclude on the effect of differential rotation on both the coupling ratios and the amplitude ratios. For the case of model~B, where meridional flow and rotation are present, the real and imaginary part of the coupling ratios and the amplitude ratios exhibit non-symmetric patterns as a function of azimuthal order $m$. Comparing this to the ratios obtained by using model~A, the effect of differential rotation is more pronounced in the real part than in the respective imaginary part. Depending on order $m$, the relative deviation of the real part of the amplitude ratios due to differential rotation exceeds 100\%, while for the imaginary part it is less than 40\% for all $m$. 

\subsection{Comparison self- and cross-coupling due to rotation}
\begin{figure}[htp]
\begin{center}
\includegraphics[width=17cm]{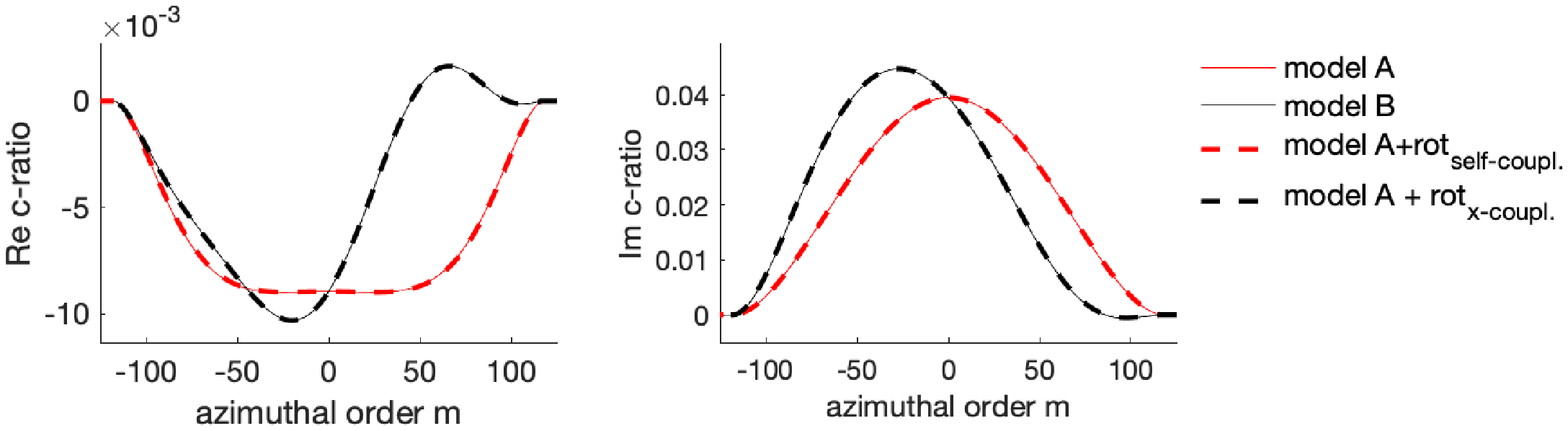}
\caption{Real part (left) and imaginary part (right) for comparison of coupling ratios calculated for the multi-cellular meridional flow model and the rotation model of Fig.~\ref{fig:rotprofiles_simu} for the example modes $(n=2,l=120)$ and $(n'=2,l'=116)$. The coupling ratios for the superposition of meridional flow and all contributions by rotation (black line) and for the superposition of meridional flow with rotation perturbations from cross-coupling only (model A+ $\text{rot}_{\text{x-coupl.}}$, black dashed line) as well as the ratios from perturbations by rotational self-coupling (model A +$\text{rot}_{\text{self-coupl.}}$, {red} dashed line) and meridional flow without rotation ({red} line) are visually not distinguishable, respectively.}
\label{fig:cratiocomp}
\end{center}
\end{figure}

We investigate the effect of self- and cross-coupling due to rotation on the coupling ratios. For comparision we show in Fig.~\ref{fig:cratiocomp} coupling ratios $C_{ik}=c_{ik}/c_{kk}$ as an example for the mode $(n=2,l=120)$ computed numerically for the three cases: \newline
a) contribution of rotation to self-coupling only, i.e., $H^{(rot)}_{k'k}=0$ for all $k'\neq k$,\newline
b) contribution of rotation via cross-coupling only, i.e., $H^{(rot)}_{kk}=0$, \newline
c) contribution of rotation to both self-coupling and cross-coupling.

Even though the matrix elements due to self-coupling are large compared to the matrix elements of cross-coupling, the obtained results clearly demonstrate that the effect of self-coupling, i.e., $H^{(rot)}_{kk}$, to the coupling ratios is negligible since the inclusion or exclusion of these terms in the computation of the coupling ratios has no influence on them. As a consequence, up to first order, only the cross-coupling of modes due to differential rotation is relevant for the perturbation of the mode eigenfunctions and thus for the amplitude ratios. This, however, is in contrast to the perturbation of the mode eigenfrequencies, where the matrix elements due to rotational self-coupling enter in first order into the perturbation analysis, cf.~\eqref{eq:ceq}, and result in the known rotational splitting of the mulitplet frequency with frequency shifts in the order of $1\,\mu$Hz. 

\subsection{Compensation of the effect of differential rotation}
According to the results obtained in \ref{subsec:cratarat} the effect of differential rotation cannot be neglected when the meridional flow is to be inferred from amplitude ratios since both perturbations superimpose. 

\begin{figure}[htp]
\begin{center}
\subfigure[$l'=116$]{
\includegraphics[width=17cm]{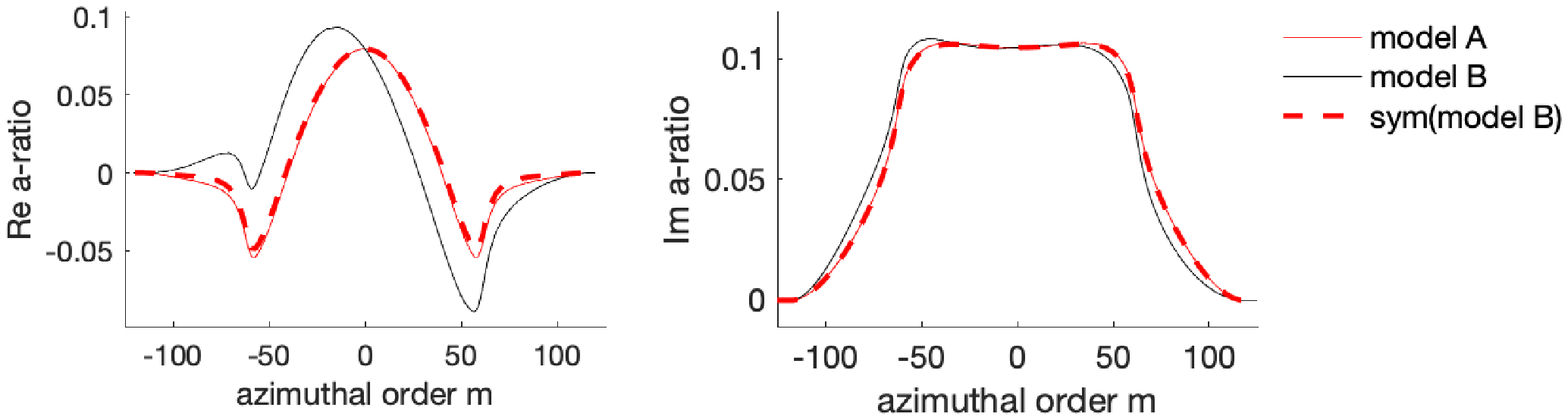}}
\subfigure[$l'=117$]{\includegraphics[width=17cm]{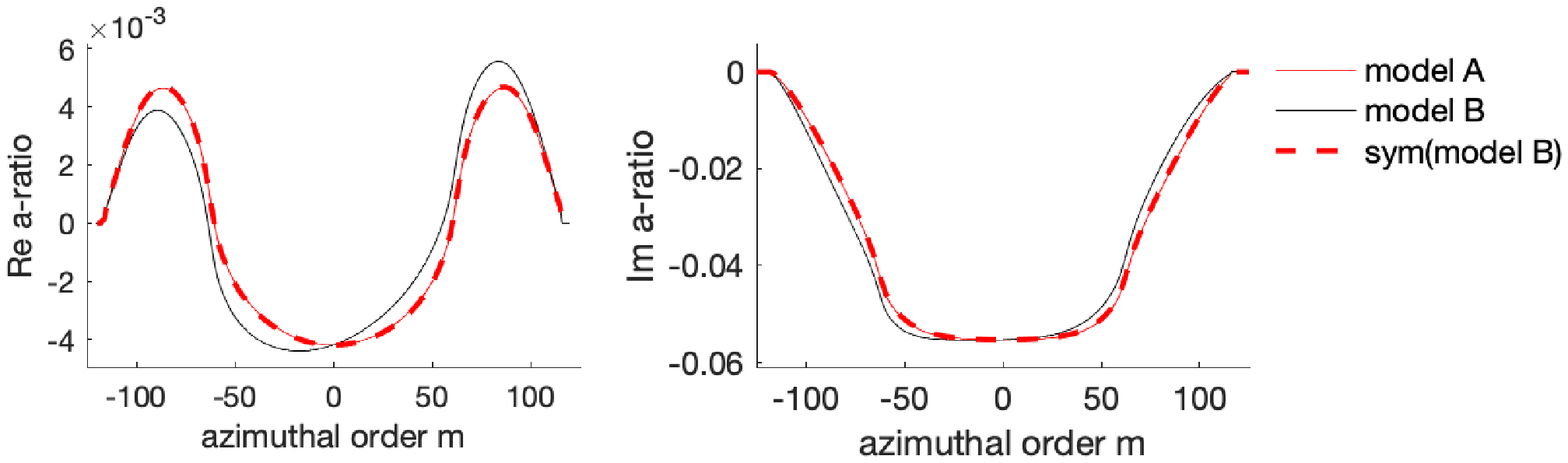}}
\subfigure[$l'=118$]{\includegraphics[width=17cm]{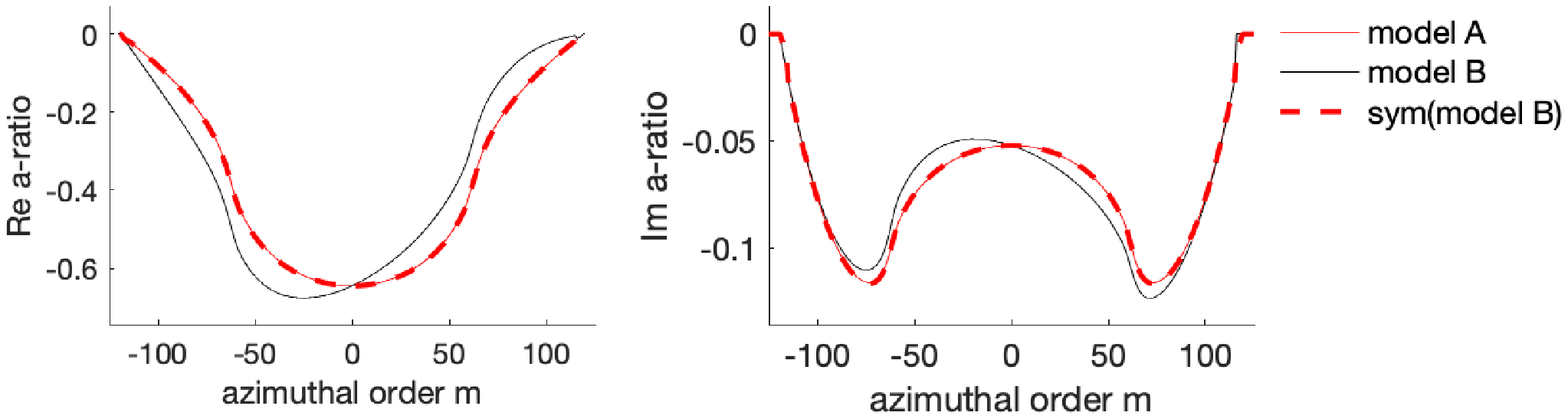}}
\caption{Profiles of amplitude ratio calculated for the meridional flow model (model A, {red} line), for the meridional flow model plus differential rotation (model B, black line), and the azimuthally symmetrized amplitude ratios (sym(model B), red dashed line) for the example mode $k=(n=2,l=120)$ and coupling modes $k'=(n=2,l')$ with: a) $l'=116$, b) $l'=117$ and, c) $l'=118$. The match between the symmetrized ratios (red dashed lines) and the ratios for the meridional flow model A is almost perfect.}
\label{fig:aratioallflows_sym}
\end{center}
\end{figure}

In~\cite{schad13} we suggested to make use of the dependency of both perturbations on the azimuthal order, i.e., symmetry in azimuthal order for the effect the meridional flow and the anti-symmetry in azimuthal order for the effect of differential rotation, in order to disentangle the respective contributions to $y_{kk'}(m)$. The azimuthal antisymmetric contribution of the differential rotation can be compensated by symmetrizing $y_{kk'}(m)$ with respect to azimuthal order $m$, i.e., 
\begin{align}
y^{sym}_{kk'}(m)=\frac{1}{2}\big(y_{kk'}(m)+y_{kk'}(-m)\big)\, .
\end{align}
Formally, this linear operation does not perfectly disentangle contributions from rotation and meridional flow in the amplitude ratios since they depend nonlinearly on the coupling ratios of rotation and meridional flow. However, this aspect is considered to be negligible as illustrated in Fig.~\ref{fig:aratioallflows_sym}. There, symmetrized amplitude ratios of a model incorporating rotation and meridional flow are compared with the respective non-symmetrized amplitude ratios and with amplitude ratios obtained from a model without the influence of solar rotation for the example mode $(n=2, l=120)$. The leakage matrix is that of the MDI instrument which is assumed to be symmetric. The symmetrized amplitude ratios are nearly identical to the amplitude ratios obtained when calculating the effect of meridional flow alone. Only minor deviations are obtained for the real part of the amplitude ratios between $l=120$ and $l'=116$ at $|m|>30$. Hence, for the presented example, the signal of differential rotation in the symmetrized amplitude ratios is almost completely compensated. The deviations are small compared to the statistical uncertainty obtained for amplitude ratios from observed data. 

\section{Data Analysis}
\subsection{Data analysis}
We analyzed splitting coefficients and gap-filled time series of spherical harmonically decomposed Dopplergrams~\citep{larson15} from the medium-$l$ structure program of the MDI instrument~\citep{scherrer95} and the HMI instrument~\citep{schou12}. The MDI data cover the period from 2004.01.08--2010-09.20, the HMI data cover the period from 2010.04.30--2014.04.08. The time series of each dataset are divided in non-overlapping segments of approximately 360d length in order to compensate for possible periodic annual variations of the systematic spatial leakage, e.g., due to variations of the $B_{0}$-angle~\citep{zaatri06}. For the MDI data we selected six segments with duty cycles $>0.95$, the HMI data are divided into four segments. A table of the selected data is given in Tab.~\ref{tab:hmi_timeperiods}.

We evaluate modes of harmonic degrees $0\leq l\leq 198$ and frequencies $1.32\,\mathrm{mHz}\leq \nu_{nl}\leq 4.77\,\mathrm{mHz}$. For each mode and data segment, the amplitude ratios and errors are estimated according to the procedure described in~\citet{schad13} (equation (7) and (8), there). In total, 12925 pairs of coupling multiplets $(k=(n,l),k'=(n',l'))$ with harmonic separation $dl=|l'-l| \leq 3$ and frequency separation $|\nu_{k}-\nu_{k'}|\leq \delta \nu= 31\,\mu$Hz are analyzed. The amplitude ratios are averaged over the six segments.

$a$-coefficients are estimated from the real part of the amplitude ratios on the basis of equation~\eqref{eq:crotmerid} and \eqref{eq:aratio} by minimization of the non-linear, weighted least-squares function. The weighting is given by the estimation error of the amplitude ratios. The parameter estimation incorporates the leakage matrix of the respective instrument. The horizontal component of the leakage matrix is neglected in the analysis, since its contribution to equation~\eqref{eq:aratio} is small compared to the radial component for modes of low and medium degree $l$. \\

Toroidal flow coefficients of degree $s=2,\,\hdots,8$ are estimated from the $a$-coefficients, see equation~\eqref{eq:acoeff},  
by means of the Subtractive Optimally Localized Averages (SOLA) inversion approach~\citep{pijpers94}. Toroidal flow kernels are computed from the eigenfunctions of solar Model S~\citep{dalsgaard96} following Eq.~\eqref{eq:torkernel}. The inversion is carried out on a grid of target positions $\{r_{j}\}_{j=1,\hdots,J}$ within the range $0.57 < r_{j}/R < 0.992$, where $R$ is the solar photospheric radius of Model S. For each degree $s$, the regularization parameter entering the SOLA inversion is adjusted to obtain flow estimates as deep as possible on the one hand and well localized inversion kernels on the other hand. The radial position and resolution of flow estimates is determined by the center of mass and standard deviation of the Gaussian shaped inversion kernels, respectively~\citep{dalsgaard90}. \\

The error of the toroidal flow coefficients is derived from the diagonal elements of the inverse Hessian matrix of the least-squares fit at the optimum and the error propagation of the SOLA inversion analysis, where we take into account the non-uniform estimation errors of the amplitude ratios. The rotation rate is estimated from the toroidal flow components according to Eq.~\ref{eq:Omega}. The error of the rotation rate is derived according to the Gaussian error propagation law.

\section{Results}
\label{sec:results}
\subsection{Inversion for Solar Rotation}
\begin{figure}[htbp]
\begin{center}
\includegraphics[width=12cm]{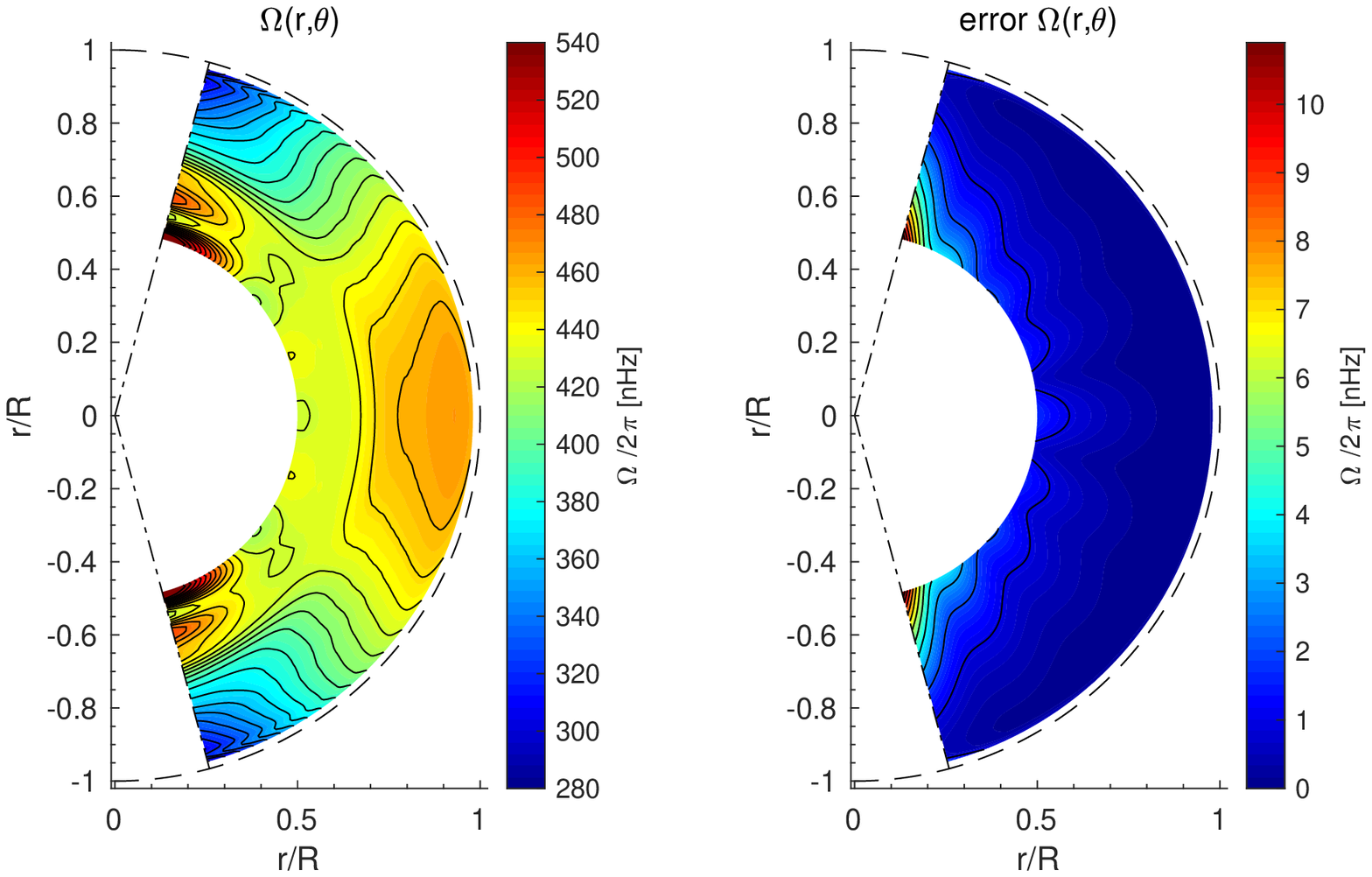}
\includegraphics[width=12cm]{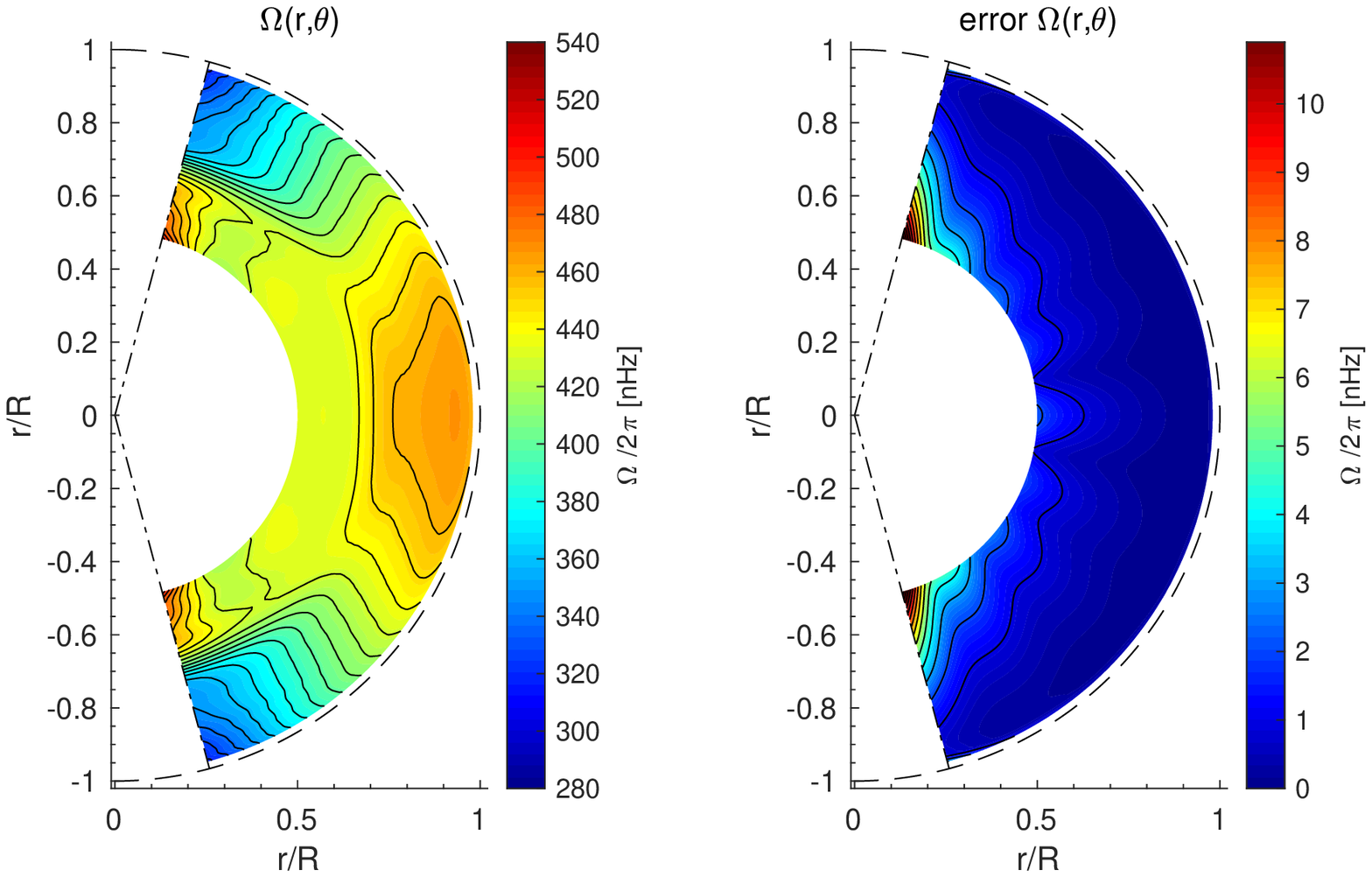}
\caption{{Left:} Rotation rate {profiles estimated from frequency-splittings from MDI data (top) and HMI data(bottom)} with the SOLA method. {Contour lines are given in increments of 10\, nHz between 280 and 540\,nHz. Right: 1$\sigma$ errors on the rotation rate estimates with contours given in increments of 1\,nHz between 0 and 15\,nHz.}}
\label{fig:rotprofilemdisplits}
\end{center}
\end{figure}

We used the SOLA method~\citep{pijpers94} for inversion of the splitting coefficients. Splitting coefficients for $k=1,\dots,36$ are taken into account. The composite rotation rate estimated from MDI data is depicted in Fig.~\ref{fig:rotprofilemdisplits}. 
\\

\begin{figure}[htp]
\begin{center}
\includegraphics[width=12cm]{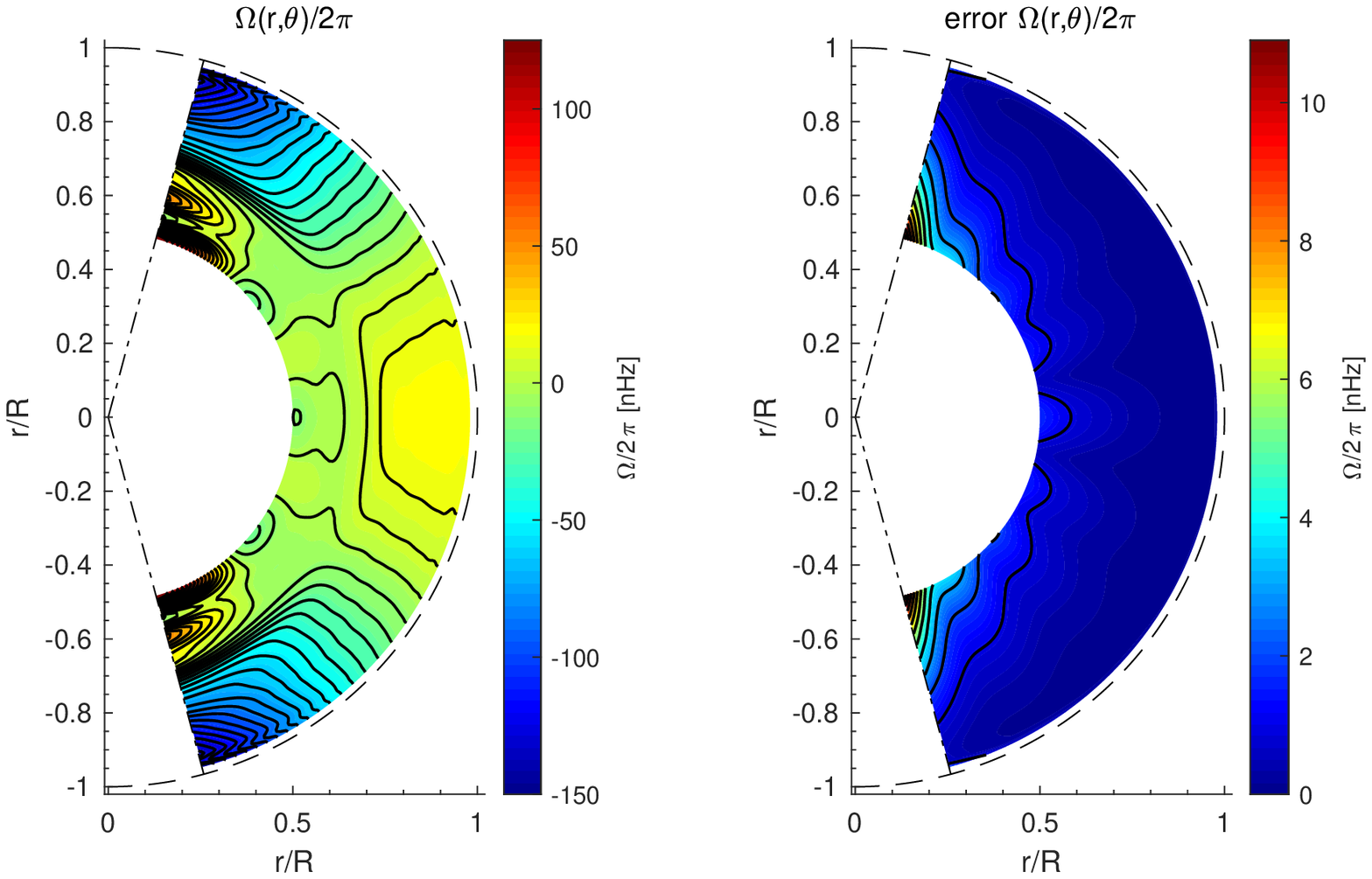}
\includegraphics[width=12cm]{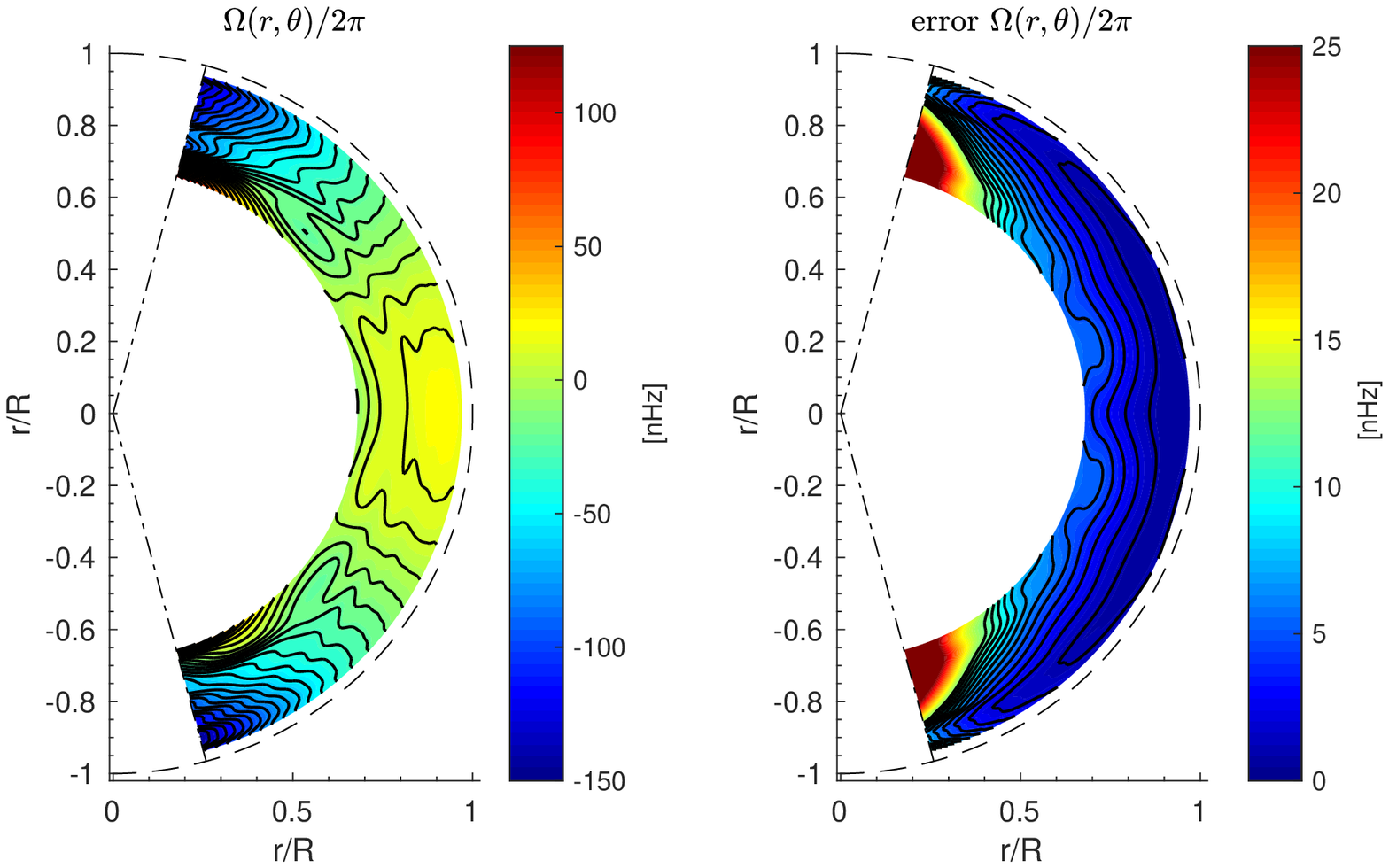}
\includegraphics[width=12cm]{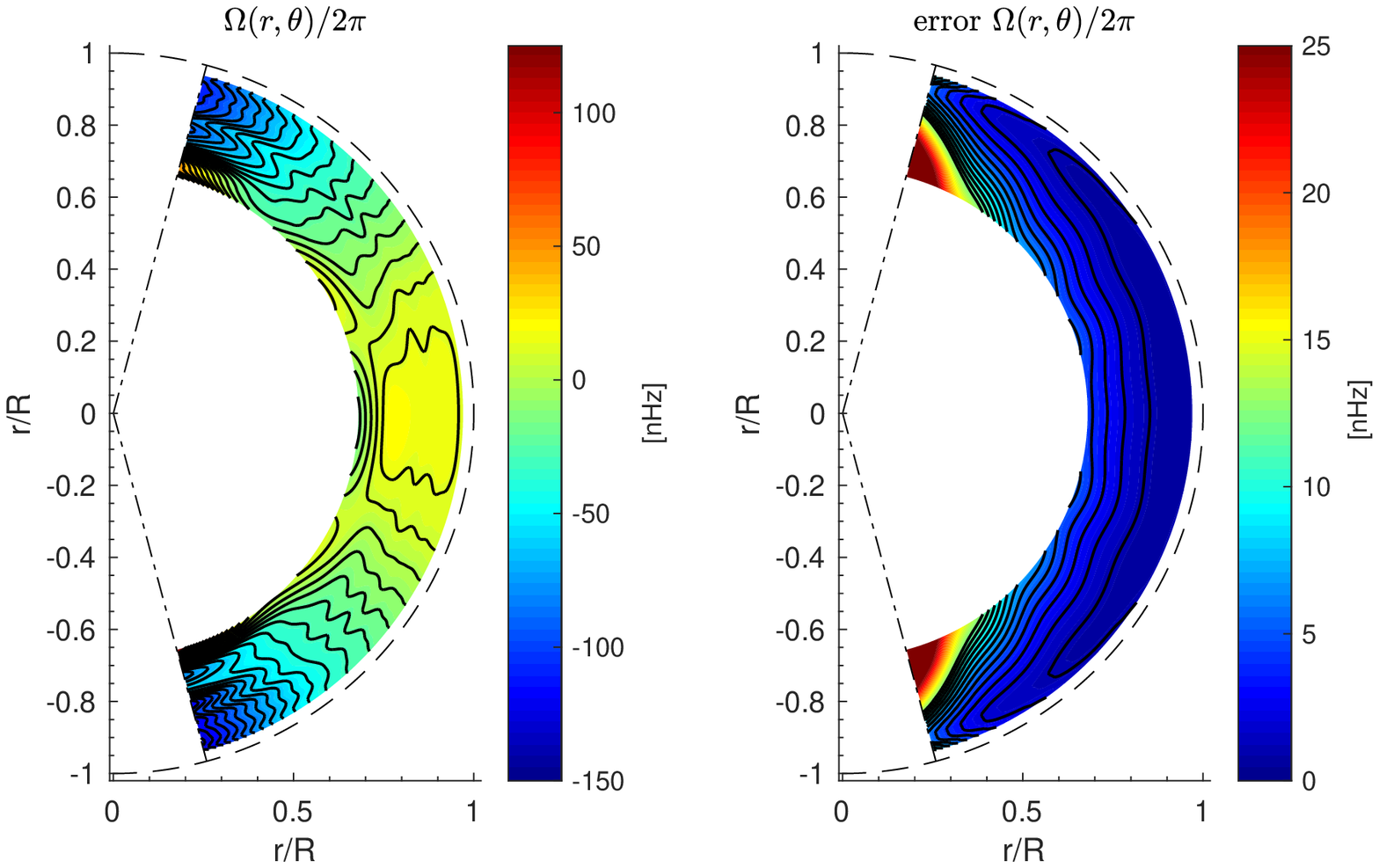}
\caption{Left: Rotation rate profile estimated from frequency-splittings from MDI data without the $s=1$ component (upper panel) and from {EFPA} $a$-coefficients from the same data set (middle panel) and from {EFPA} $a$-coefficients from HMI data (bottom panel). {Note, the HMI data does not cover the same period as the MDI data set.} Right: $1\sigma$ error of {the} estimated rotation rates on the {left} side.}
\label{fig:rotmdi}
\end{center}
\end{figure}

\begin{figure}[htp]
	\begin{center}
		\includegraphics[width=17cm]{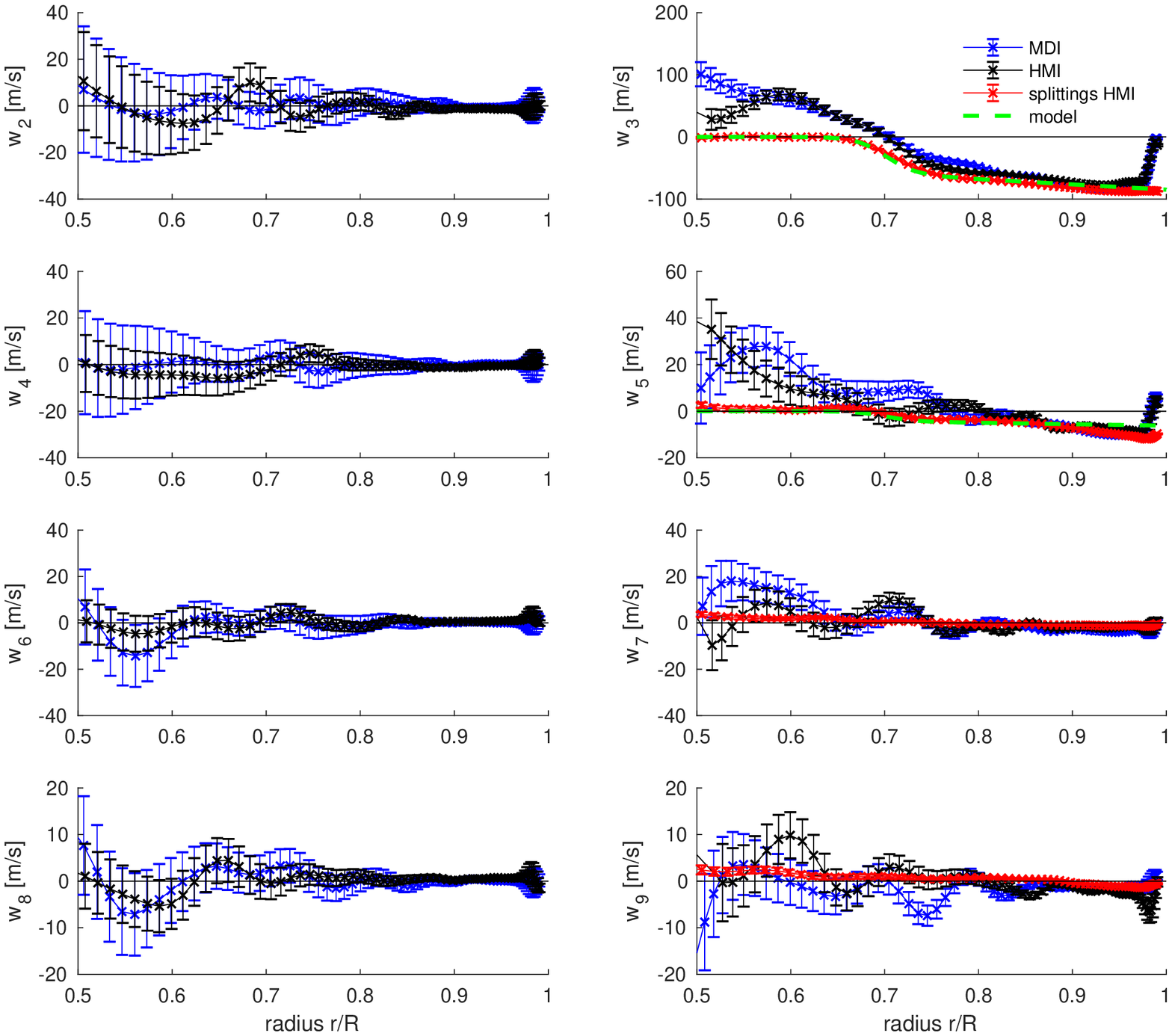}
		\caption{Rotation rate coefficients $w_{s}$ for $s=2,\dots,9$ as a function or $r/R$ estimated from frequency splittings of HMI data (red crosses) and $a$-coefficients from MDI (blue crosses) and HMI (black crosses) data together with their $1\sigma$ standard error. A green dashed line depicts the $w_{3}$- and $w_{5}$-coefficient from the differential rotation model in Section~\ref{diffmodel}.} 
		\label{fig:wscoeffs}
	\end{center}
\end{figure}

The inversion of $a$-coefficients is carried out analogously as of the $b$-coefficients for the meridional flow in~\cite{schad13}. Rotation rate profiles without the first component $w_{1}$ estimated from MDI data via splitting coefficients and the $a$-coefficients from both, MDI and HMI data, are depicted in Fig.~\ref{fig:rotmdi}. 
Individual toroidal flow coefficients $w_{s}(r)$ estimated from $a$-coefficients of MDI and HMI data in comparison with those obtained from HMI frequency splittings are depicted in Fig.~\ref{fig:wscoeffs} as a function of radius. 

On a first view, we note that inversions of frequency splittings and $a$-coefficients are in qualitative good agreement.

\begin{figure}[htbp]
\begin{center}
\includegraphics[width = 12cm]{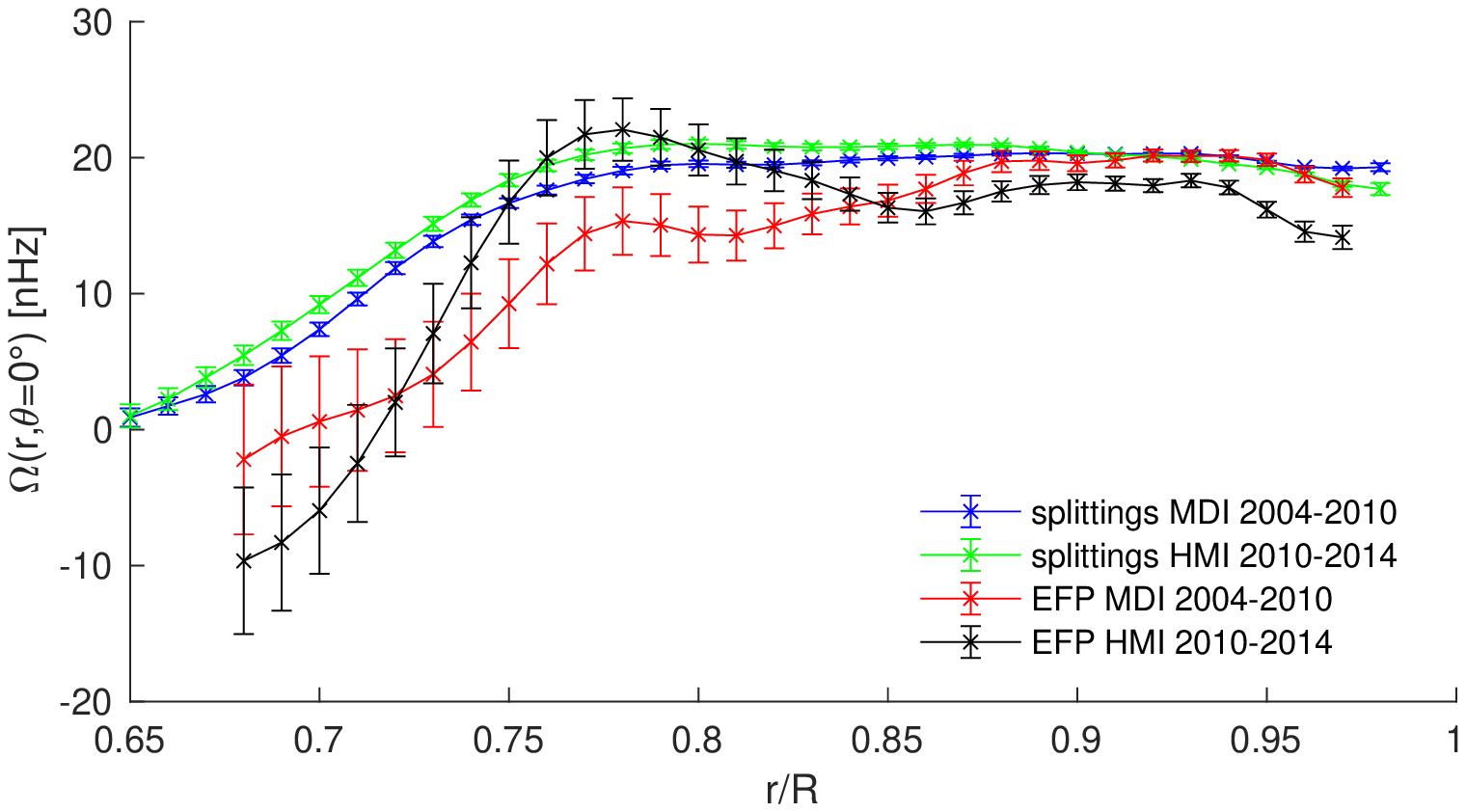}
\caption{Rotation rates at the solar equator estimated from frequency-splittings and $a$-coefficients from MDI and HMI data as a function of radius $r/R$ together with their $1\sigma$ error bar.}
\label{fig:rotmdivshmi}
\end{center}
\end{figure}

\begin{figure}[htp]
\begin{center}
\includegraphics[width=12cm]{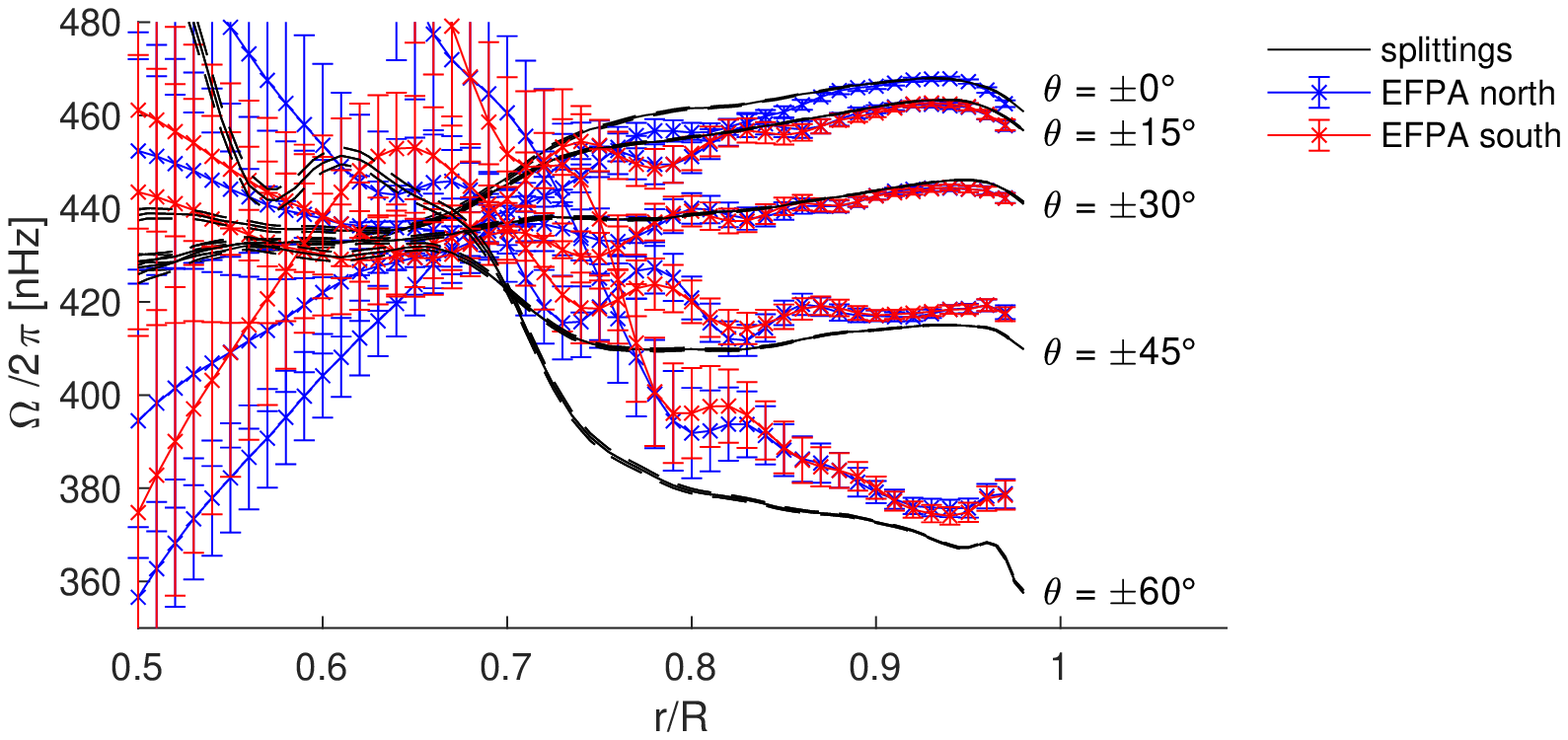}
\includegraphics[width=12cm]{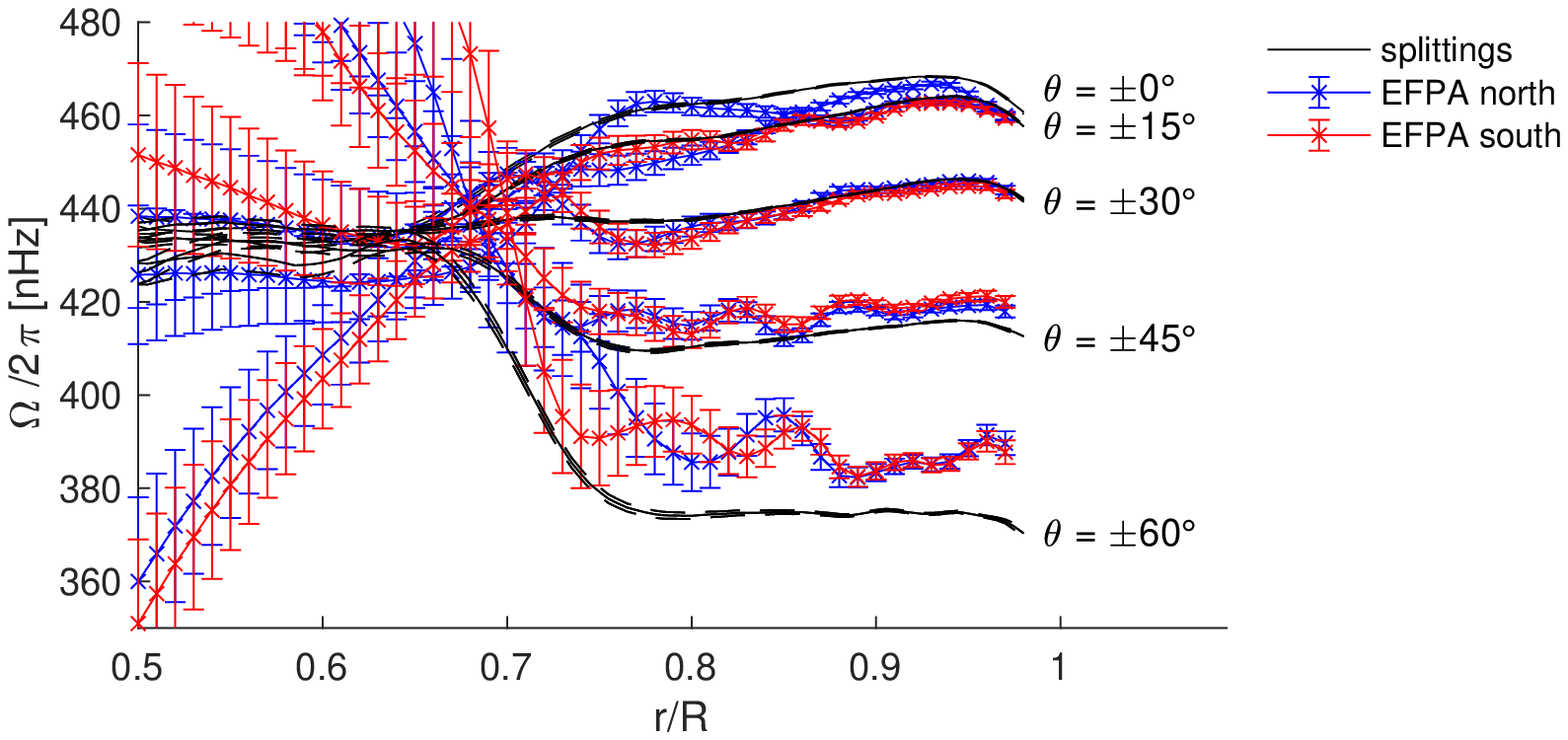}
\caption{Comparison of rotation rate profiles reconstructed from frequency splittings and $a$-coefficients estimated from MDI { (top) and HMI (bottom)} data as a function of $r/R$ at distinct latitudes between $\theta=0,\dots,\pm 60$ degree. {The rotation rates obtained from frequency splittings are averages over North and South and given in black, while the ones from EFPA are given in blue for the North and in red for the South.}}
\label{fig:comparison}
\end{center}
\end{figure}

\begin{figure}[htp]
	\begin{center}
		\includegraphics[width=12cm]{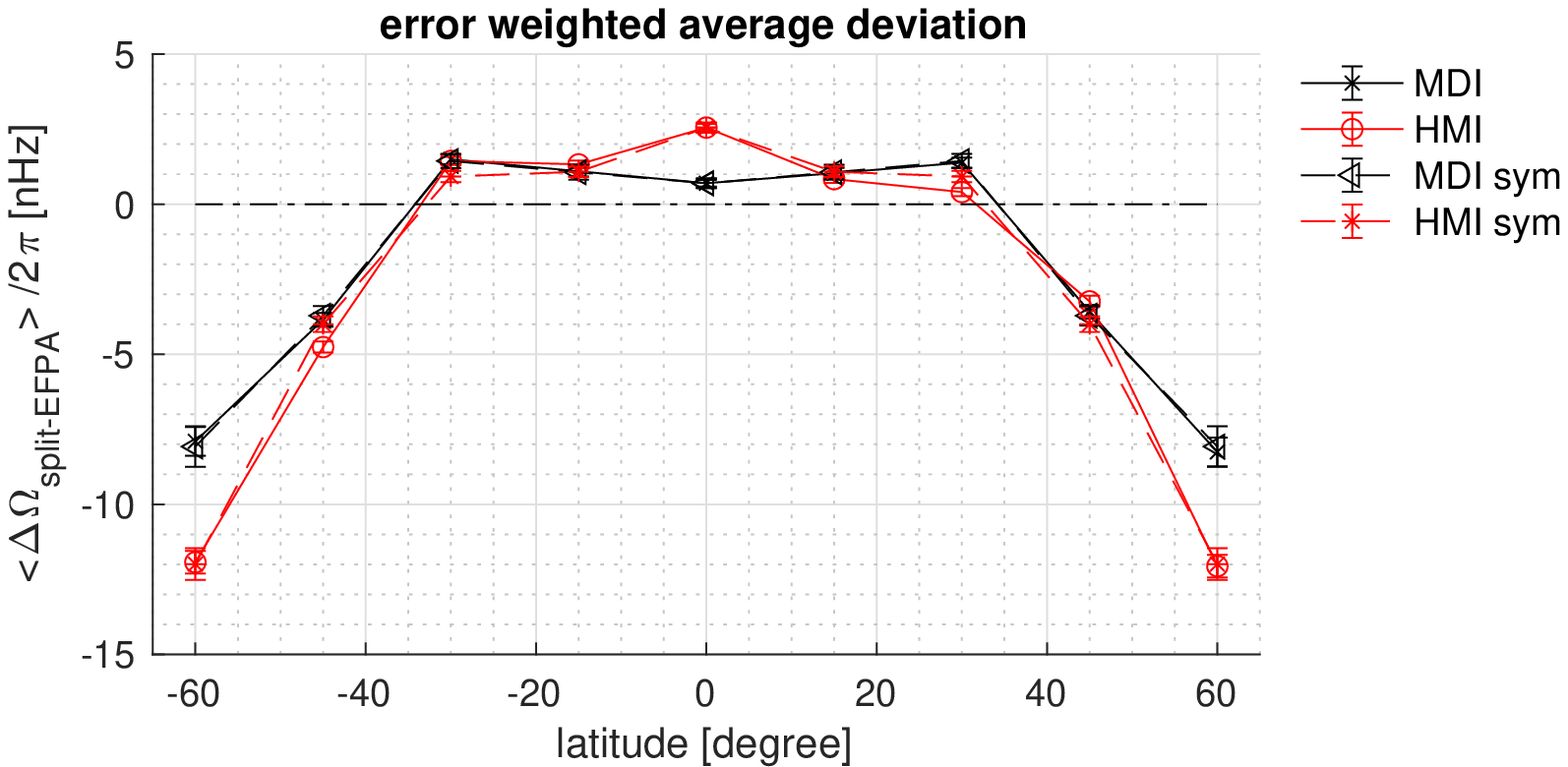}
		\caption{Error-weighted average deviations between the obtained rotation rates by EFPA and those from the frequency splittings. The deviations are shown as a function of latitude for MDI (black) and HMI (red). The averaging was carried out over the depths 0.73 to 0.97~R. Solid lines give the deviation of the asymmetric EFPA results from the respective frequency splitting inversions, while the dashed lines show the deviations after averaging the EFPA results over North and South.}
		\label{fig:avgcomp}
	\end{center}
\end{figure}

Investigating the details, there are significant differences between the rotation rate profiles from different estimation methods, instruments and different observation periods. The errors on the EFPA inversions are naturally higher. The inversion of the frequency splittings, however, is only able to provide an averaged estimate of the rotation rate over both hemispheres. The EFPA inversion delivers information on the rotation rate for the northern and southern hemispheres independently, indicating asymmetries in the rotational rates between North and South.

This is further illustrated in Fig.~\ref{fig:rotmdivshmi}. At the solar equator, the rotation rate profiles from the MDI data from splittings and $a$-coefficients are in good agreement in the upper part of the convection zone, i.e. for $r\gtrsim$0.85~R. Below 0.85~R, the rotation rate from EFPA decreases significantly stronger with increasing depth compared to the rotation rate from splittings. The rotation rate profiles from splittings between MDI and HMI do not show significant deviations of each other except near the surface. However, the HMI rotation rate profile from EFPA is significantly smaller above $\gtrsim$0.82~R. 
Interestingly, although accompanied by a large error, the rotation rate of the EFPA for MDI and HMI data are crossing zero at about 0.73~R. This is not the case for the profiles from splittings, which show a deeper reaching latitudinal differential rotation profile getting zero at about 0.65~R. 

{
Fig.~\ref{fig:comparison} provides a quantitative comparison of the rotation rate at latitudes between {$0^\circ$ and $\pm 60^\circ$} derived from the EFPA method and the frequency splittings for MDI and HMI. The latter is an average over North and South. 
A similar behaviour as shown in Fig.~\ref{fig:rotmdivshmi} can be found for the near-equatorial latitudes up to $30^\circ$, too: The rotation rate from EFPA is slightly lower than that of the classical approach. Overall, there is good agreement between the inversions obtained from frequency splittings and the EFPA inversions throughout the convection zone. 
For $\theta\ge 45^\circ$, EFPA gives higher estimates for the rotation rate then the splitting coefficients. 
As already noted above, the EFPA method has larger error bars than the classical approach. Taking this into account and considering the results of the frequency splittings as reference, the EFPA method delivers reliable results down to approximately 0.75~R. }

{There is an indication of larger deviations from the frequency splitting results and from symmetry in HMI data than for MDI data, which is implied in Fig.~\ref{fig:rotmdi} and in Fig.~\ref{fig:comparison}. This is further investigated with the help of Fig.~\ref{fig:avgcomp}. There, the difference between the rotation profiles obtained from frequency splittings and EFPA in the depth range 0.73 -- 0.97~R is used to create an error-weighted average deviation for each latitude. 	

We find that HMI shows stronger deviations from frequency splitting results at higher latitudes and at the equator than MDI. Subtracting the deviations of the North from that of the South and summing up the absolute values of these differences, results in a symmetry score value. With $3.2\pm 0.7$ nHz this score value is significantly higher for HMI than for MDI, which has a score value of $0.8\pm 0.8$. Averaging first the EFPA results over North and South, and then determining the error-weighted average deviation from the frequency splittings is also shown in Fig.~\ref{fig:avgcomp}. In this case, the average deviations are reduced at maximum by 0.8~nHz for HMI and 0.2~nHz for MDI, resp.} 

\section{Discussion and conclusion} 
\label{sec:discussion}
In this paper we considered the influence of rotation on p-mode eigenfunctions and adapted the EFPA method used previously for estimation of the meridional flow to infer rotation from mode eigenfunction perturbations. We estimated rotation rate profiles from data from the MDI and HMI instrument and compared the results to conventional global helioseismic rotation inferred from frequency splittings. The effect of rotation is clearly observable in the amplitude ratios as illustrated by a computational model. 

The perturbation of eigenfunctions due to rotation is, unlike to meridional flow, characterized by real valued coupling coefficients with an azimuthally antisymmetric signature. This has consequences on the measurable amplitude ratios. As shown by numerical analyses with different flow models, the real part of the amplitude ratios is essentially influenced by instrumental leakage and solar rotation, whereas the imaginary part of the amplitude ratios is mainly determined by perturbations of first order due to the meridional flow. Perturbative influences from the meridional flow of higher orders can also contribute to the real part of the amplitude ratio, but they are expected to be small compared to leakage and rotation. 

For the here examined rotation model and modes we found that the influence of rotational self-coupling on the coupling ratios and amplitude ratios is negligible since self-coupling only contributes to the higher order terms of the coupling ratio although the matrix elements of rotational cross-coupling are much smaller in magnitude than the respective ones for self-coupling. Cross-coupling terms have a crucial influence on the amplitude ratios. This result reflects that only deviations of the supermatrix from a diagonal matrix are able to mix the unperturbed eigenfunctions. 

Leakage and the meridional flow both feature an azimuthal symmetric signature, whereas rotation is characterized by an azimuthal antisymmetric signature. These distinctive symmetry properties can be utilized to compensate for the influence of rotation on the amplitude ratio by symmetrization of the amplitude ratios. This processing is not exact, but we have shown with computational models that symmetrization of amplitude ratios with respect to order $m$ is able to compensate for azimuthal antisymmetric influences of differential rotation on the amplitude ratios to a good degree of accuracy compared to estimation errors. This processing step is of importance for inferences on the meridional flow from amplitude ratios and used in~\cite{schad12,schad13}. Only a small bias is expected between the symmetrized amplitude ratio incorporating rotation and the amplitude ratio in the absence of rotation.

An interesting and significant property of the EFPA for rotation is its sensitivity to equatorial non-symmetric solar rotation components. This is in contrast to the frequency splittings which are not sensitive for these components at all. So far, deviations of the solar rotation rate from symmetry were only measurable by local helioseismic methods which are usually not sensitive to the deeper solar interior. \\

An important issue in current helioseismic investigations is the validation of the various helioseismic methods and results using either artificial data or by comparative studies. Here we compared two different global helioseismic approaches applied to solar data. 

We find significant differences especially at {higher latitudes and towards} the surface between the results from both methods. Their origin is so far unclear. The discrepancies might indicate systematic contributions as well as different sensitivities of the mode properties on the rotation at higher latitudes. Interestingly, the results from the MDI {data show} a better agreement as the ones from the HMI data. {We conclude that knowledge on the leakage matrix has to be as good as possible to reduce systematic effects in the EFPA method. A possible consideration of the effect of $B$ angle variations might reduce the discrepancies at the higher latitudes and the greater depths.}

So far we can only speculate about this observation. The MDI data cover the minimum phase of cycle 23 while the HMI data are acquired during the rising activity phase of cycle 24. The change in the distortion of $p$ modes by the increased magnetic activity could be {another} source for this observation. This assumption needs further proof. Masking activity regions is a common procedure in local helioseismology to reduce influences from strong, local magnetic fields, but this method is considered to be not appropriate for this kind of global helioseismic investigation. Additionally, it is an extremely costly procedure in our case, since this would also affect the leakage matrix, which would need to be modified according to the masking. The further analyses about the origin of this discrepancy will be subject of a future study which may help to give deeper insights into sensitivities and systematic effects, e.g. by magnetic fields, on helioseismic inversions.

\subsection*{Acknowledgements}
{The authors thank Tim Larson and Jesper Schou for their support in obtaining MDI and HMI data.}
The theoretical and simulation part of the presented work was supported by the Deutsche Forschungsgemeinschaft DFG, Grant No. Ti315/4-2. For the analysis of HMI and MDI data, A.S. and M. R. received funding from the European Research Council under the European Union's Seventh Framework Program (FP/2007 2013)/ERC grant Agreement No. 307117. {The authors thank the anonymous referee for their useful comments on the manuscript.}

\appendix
\section{Toroidal flow kernel}
The toroidal flow kernel for solar rotation from~\cite{ritzwoller91,lavely92} is

\begin{align}
\label{eq:toroidalkernel}
T^{k'k}_{s}=&-\frac{1}{2}(1-(-1)^{l'+s+l})\sqrt{l'(l'+1)l(l+1)}\begin{pmatrix}
     l' &  s & l  \\
     -1 & 0  & 1\\
\end{pmatrix}\notag\\
&\times\frac{1}{r}\,\Big\{\xi_{k'}^{r}\xi_{k}^{h}+\xi_{k'}^{h}\xi_{k}^{r}-\xi_{k'}^{r}\xi_{k}^{r}+\frac{1}{2}\xi_{k'}^{h}\xi_{k}^{h}\left[s(s+1)-l(l+1)-l'(l'+1)\right]\Big\}\, ,
\end{align}
where $T^{k'k}_{s}=T^{kk'}_{s}$, and $T_{s}^{k'k}=0$ if $(l'+s+l)$ is even. The kernel is composed of two contributions that might be loosely denoted as the advection contribution $\xi_{k'}^{r}\xi_{k}^{r}+1/2[l(l+1)+l'(l'+1)]\xi_{k'}^{h}\xi_{k}^{h}$ and the Coriolis contribution $\xi_{k'}^{r}\xi_{k}^{h}+\xi_{k'}^{h}\xi_{k}^{r}+1/2 s(s+1)\xi_{k'}^{h}\xi_{k}^{h}$~\citep{ritzwoller91}. We note that the sign of the kernel in Eq.~\eqref{eq:toroidalkernel} is reversed to the formulation presented in~\cite{ritzwoller91,lavely92} since the time dependency there is given by $\exp(\I \omega t)$ instead of $\exp(-\I \omega t)$ as used in this study.
The matrix element, defined similar to ~\cite{schad11}, is
\begin{align}
H^{k'k}=8\pi \omega_{ref}(-1)^{m'}\gamma_{l}\gamma_{l'}\sum_{s}\gamma_{s}\begin{pmatrix}
     l' &  s & l  \\
     -m' & 0  & m\\
\end{pmatrix}\,\int_{0}^{R}\rho_{0}(r)w_{s}(r)T_{s}^{k'k}(r)r^{2}dr
\end{align}
where $\gamma_{l}=\sqrt{(2l+1)/(4\pi)}$.
\section{Relation of Toroidal Flow Coefficients and Rotation Rate}
\label{app:rotationws}
The first three non-vanishing toroidal expansion coefficients $w^{0}_{s}(r)$are related to $\{\Omega_{k}(r)\}_{k}$ by~\citep{ritzwoller91} 
\begin{subequations}
\begin{align}
w_{1}^{0}(r)&=2\sqrt{\frac{\pi}{3}}\,r\left[\Omega_{0}(r)-\frac{1}{5}\Omega_{2}(r)\right]\, , \label{eq:w1}\\
w_{3}^{0}(r)&=2\sqrt{\frac{\pi}{7}}\,r\left[\frac{1}{5}\Omega_{2}(r)-\frac{1}{9}\Omega_{4}(r)\right]\, , \label{eq:w2}\\
w_{5}^{0}(r)&=2\sqrt{\frac{\pi}{11}}\,r\left[\frac{1}{9}\Omega_{4}(r)\right]\, , \label{eq:w3}
\end{align}
\end{subequations}
where contributions from $\Omega_{k>4}$ are ignored.  
\section{Meridional Flow Model}
\label{sec:simumflow}
For our simulation study we use a numerical model of the meridional flow $\mbf{u}(r,\theta)$, which consists of closed poloidal flow cells of various harmonic degrees $s$. These cells are confined between the solar surface $R$ and the bottom of the convection zone $r_{b}=0.713\,$R~\citep{basu97}. These meridional flow components are completely determined by the specification of the radial flow strengths $u^{0}_{s}(r)$, which is parameterized analogously to~\cite{roth08} by
\begin{eqnarray}
\label{eq:usmodel}
u_{s}^{0}(r)=\left\{\begin{array}{ll}
A_{s} \sin\big(n_{s}\pi \frac{r-r_{b}}{R-r_{b}}\big)  & r_{b}\leq r \leq R \\
0 & \mathrm{otherwise} \, .
\end{array}\right.
\end{eqnarray}
Here $n_{s}$ is the number of flow cells in depth, $s$ defines the number of poloidal flow cells in latitude, and $A_{s}$ determines the maximum flow strength. The radial flow components vanish at the bottom and top of the confining shell, i.e., $u_{s}^{0}(R)=u_{s}^{0}(r_{b})=0$ for all degrees $s$. The horizontal velocity components are determined by the radial flow components via mass conservation expressed as~$\rho_{0}rs(s+1)v_{s}^{0}=\partial_{r}(r^{2}\rho_{0}u_{s}^{0})$. Following this relation and Eq.~\eqref{eq:usmodel} one finds for the horizontal flow strength at the surface and at the bottom of the convection zone:
\begin{subequations}
\begin{align}
\label{eq:vsR}
v^{0}_{s}(R)&= (-1)^{n_{s}}\frac{n_{s}\pi}{s(s+1)} \frac{R}{R-r_{b}} A_{s}\\
v^{0}_{s}(r_{b})&= \frac{n_{s}\pi}{s(s+1)} \frac{r_{b}}{R-r_{b}} A_{s}=(-1)^{n_{s}} \frac{r_{b}}{R}v^{0}_{s}(R)\, . 
\end{align}
\end{subequations}

This meridional flow model is composed of multiple flow components with degrees $s=1,\dots, 5$ with varying number of flow cells in depth $n_{s}$ and different flow strengths $A_{s}$. The flow parameters are chosen arbitrary and listed in~Tab.~\ref{tab:model2}. \\

\begin{table}[ht]
\caption{Parameterization of the meridional flow model -- number of flow cells in latitude $s$, the number of flow cells in depth $n_{s}$, and the flow strength $A_{s}$.}
\begin{center}
\begin{tabular}{c|ccccc}
$s$ &1& 2 & 3 & 4 & 5 \\
\hline
$n_{s}$ & 1 & 2 & 1& 3 & 1 \\
\hline
$A_{s}$ [m/s] & 3 & 15 & -12 & 9 & -1.5 \\
\label{tab:model2}
\end{tabular}
\end{center}
\end{table}%

\begin{figure}[htp]
\centering
 \includegraphics[width=\textwidth]{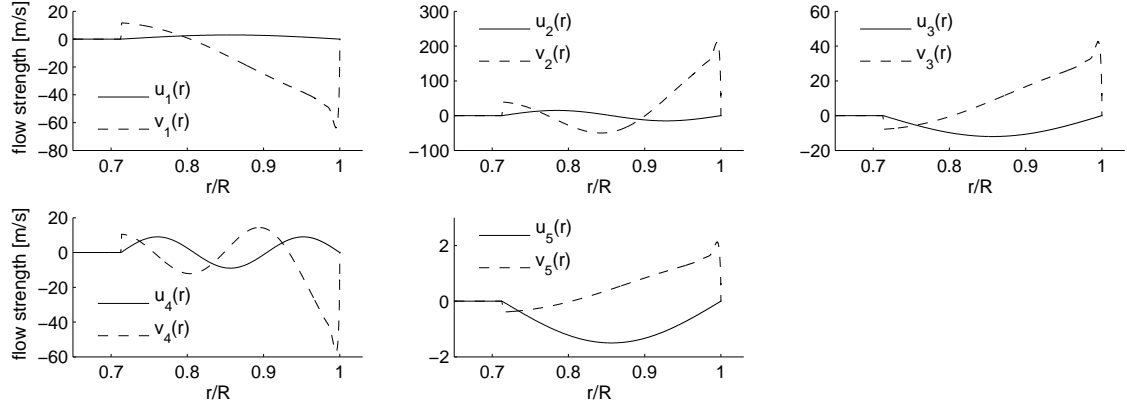}
\caption{Radial and horizontal flow profiles for the degrees $s=1,\dots,5$ of the multi-cellular meridional flow model.}
\label{fig:flowprofMod1}
\end{figure}

The radial and horizontal flow profiles of each degree $s$ are depicted in Fig.~\ref{fig:flowprofMod1}. 

\section{HMI and MDI data}
The following Table~\ref{tab:hmi_timeperiods} gives date and epoch number of the 72d long MDI and HMI time series merged into segments of about 1 year length.
\begin{table}[ht]
\caption{Observation periods and T\_START index of selected MDI and HMI data.}
\begin{center}
\begin{tabular}{|c|c|c|}
\hline
Segment & T\_START index & Period\\
\hline
\textbf{MDI} & & \\
\hline1 & 4024 -- 4312 & 2004.01.08 -- 2005.01.01 \\
2 & 4384 -- 4672 & 2005.01.02 -- 2005.12.27\\
3 & 4744 -- 5032 & 2005.12.28 -- 2006.12.22\\
4 & 5104 -- 5392 & 2006.12.23 -- 2007.12.17\\
5 & 5464 -- 5824 & 2007.12.18 -- 2009.02.21\\
6 & 6112 -- 6400 & 2009.09.26 -- 2010.09.20\\
\hline
\textbf{HMI}& & \\
\hline
1 & 6328 -- 6616 &  2010.04.30 -- 2011.04.24\\
2 & 6688 -- 6976 & 2011.04.25 -- 2012.04.18\\
3 & 7048 -- 7336 & 2012.04.19 -- 2013.04.13\\
4 & 7408 -- 7696 & 2013.06.25 -- 2014.04.08\\
\hline
\end{tabular}
\end{center}
\label{tab:hmi_timeperiods}
\end{table}%

\clearpage
\bibliographystyle{plainnat}
\bibliography{schad_rot}
\end{document}